\documentclass[a4paper,10pt,onecolumn,showpacs]{revtex4}
\usepackage{amssymb}
\usepackage[pdftex]{graphicx}	
	\DeclareGraphicsExtensions{.pdf,.png,.jpg}
\usepackage[applemac]{inputenc}
\usepackage{amsmath,amssymb,amsthm}
\usepackage{mathbbol,bbm}
\usepackage{graphicx,float}
\usepackage[final]{showkeys}
\usepackage{bm,dsfont}

\renewcommand{\c}[1]{\mathcal{#1}}

\renewcommand{\r}[1]{\mathrm{#1}}

\newcommand{\calC}{\mathcal{C}}

\newcommand{\calH}{\mathcal{H}}

\def\ket#1{\left|#1\right>}
\def\bra#1{\left<#1\right|}

\begin{document}
\title{Generalized Toric Codes Coupled to Thermal Baths}
\author{O. Viyuela, A. Rivas and M.A. Martin-Delgado}
\affiliation{Departamento de F\'{\i}sica Te\'orica I, Universidad Complutense, 28040 Madrid, Spain}

\vspace{-3.5cm}

\begin{abstract}
We have studied  the dynamics of a generalized toric code based on qudits at finite temperature
by finding the master equation coupling the code's degrees of freedom to a thermal bath.
As a consequence, we find that for qutrits new types of anyons
and thermal processes appear that are forbidden for qubits. These include 
 creation, annihilation and  diffusion throughout the system code.
 It is possible to solve the master equation in a short-time regime and 
 find expressions for the decay rates as a function of the dimension $d$ of the qudits.
Although we provide an explicit proof that the system relaxes to the Gibbs state for arbitrary qudits, we also prove that above a certain crossover temperature, qutrits initial decay rate is smaller than the original case for qubits. Surprisingly this behavior only happens with qutrits and not with other qudits with $d>3$.
\end{abstract}

\pacs{03.65.Yz,03.67.Pp,03.65.Vf,75.10.Jm}
\maketitle

\section{Introduction}
\label{sec:I}

It is known that the fragility of quantum states in the presence of interaction with an environment represents the main challenge for the large scale implementation of
quantum information devices in quantum computation and communication.  Quantum error correction is the theoretical method that was devised to protect a quantum
memory or communication channel from external noise \cite{Shor1, Steane, Shor2,Kitaev,Got,CRSS,Preskill,Gottesman}.
In these quantum error correction schemes,  to improve the stability of quantum information processing, the logical qubits should  be implemented in many-particle systems, typically $N$ physical spins per logical qubit. This is the quantum version of the classical method based on encoding information by repetition or redundancy of logical bits in terms of physical bits \cite{NC, rmp}. The logical qubits should be stable objects with efficient methods of state preparation, measurements and application of gates. By efficiency we mean certain scaling behavior, e.g.\ the lifetime of a logical qubit should grow with $N$.

In order to implement fault-tolerant methods for quantum information processing, we need to find a physical system with good enough properties to accomplish this protection
from noisy environment and decoherence. One promising candidate are topological orders in strongly correlated systems. Here, the ground state is a degenerate manifold of states
whose degeneracy depends on the topological properties of a certain lattice of qubits embedded into a surface with non-trivial topology \cite{Kitaev3}.  Many-body interacting
terms in a Hamiltonian are responsible for the existence of this topological degeneracy. The logical qubits are stored in global properties of the system represented by
non-trivial homological cycles of the surface. In this topological codes, the property of locality in error detection and correction is of great importance both theoretically and for practical implementations \cite{DennisKLP2002,MA2,KDP11}.  It is also possible to generalize this topological codes for units of quantum information based on multilevel
systems known as qudits, i.e. $d$-level systems, \cite{MA1,BB07,APS09,Anderson11} and study its local stability \cite{local_stability10}.
An alternative scheme to manipulate topological quantum information is based not in the ground-state properties of the system but in its excitations \cite{Kitaev3}.
These are non-abelian anyons that can implement universal gates for quantum information \cite{Nayak}.
Yet, being within the framework of topological codes based on ground state properties, it is possible to formulate new surface codes known as topological color codes (TCCs)
 \cite{Color_Codes} such that they have enhanced quantum computational capabilities while preserving its nice locality properties \cite{Color_Codes, fowler11, SR11}.
TCCs in two-dimensional surfaces allows for the implementation
of quantum gates in the whole Clifford group. This makes possible: quantum teleportation, distillation of entanglement and dense coding in a fully topological scenario.
Moreover, with TCCs in 3D spatial manifolds is possible to implement the quantum gate $\pi/8$ thereby allowing for universal quantum computation
\cite{Universal_color_codes, D_color_codes}. Very nice applications of topological surface codes have been found in other areas \cite{N_representability11, holographies11}.

Acting externally on topological codes, in order to cure the system from external noise and decoherence, produces benefits from the locality properties of these codes.
Namely, a very important figure of merit is the error threshold of the topological code, i.e. the critical value of the external noise below which it is possible to perform
quantum operations with arbitrary accuracy and time. For toric codes with qubits, the error threshold is very good, about 11\% \cite{DennisKLP2002}. This value
is obtained by mapping the process of error correction to a classical Ising model on a 2D lattice with random bonds. Interestingly enough, this type of mapping can be
made more general and applied to TCCs yielding the same error threshold \cite{Random3body09} while maintaining its enhanced quantum capabilities
\cite{RandomUJ09, Tricolored10}. These results have been
confirmed using different types of computation methods \cite{masa09, ON09, WFHH09, FWH10, LAR11}. It is also possible to carry out certain
 computations by changing the code geometry over time,
 something called ‘code deformation’ \cite{DennisKLP2002, RHG07, code_deformation07} that allows us to perform quantum computation in a different way.
A more general type of codes can be constructed with quantum lattice gauge theories based on quantum link models \cite{wiese97}.

In this paper we adopt a different approach than external protection of topological codes. 
Hence, instead of performing active error correction, we just rely on the robustness of a Hamiltonian that has a gap above the ground state manifold where the quantum information is stored. 
Thus, we leave the system to interact with the surrounding environment and study the
fate of the topological order under these circumstances. This source of noise is inescapable: the microscopic interactions of the physical spins with thermal particles
or excitations of the local environment. 
The analogous situation for classical information processing is well-understood, but the existence of a similar mechanism for quantum information is still an open problem.
The quantum theory of open systems provides a natural framework for studying stability in the presence of thermal noise.  The particularly simple properties of Kitaev's model allow us to apply the Davies' theory, namely the dynamics of a quantum system weakly interacting with a heat bath in the Born-Markov approximation
\cite{Davies,AlickiLendi,Libro,Breuer,Frigerio,Spohn1,Spohn2,SpohnReview}.
There have also been some related studies regarding thermal effects on Adiabatic Quantum Computation \cite{Sarandy,Ashhab}.

The first indication that the toric code for qubits in 2D spatial dimensions is unstable against thermal noise was shown in \cite{DennisKLP2002}.
Further analytical and quantitative arguments of thermal instability were given by \cite{Nussinov}. 
Later, a rigorous proof of this fact has been established using the theory of quantum open systems \cite{Horodecki,Alicki-Fannes-fidelity}. Subsequently, other
investigations have been conducted for abelian models, non-abelian models, TCCs \cite{IPGAP10,IPGAP09,Kargarian09,thermal_kitaev_ladders09,Kim_thermal11} etc.
Remarkably enough, while with qubits in 2D lattice models the topological protection is lost under the action of thermal fluctuations \cite{BT09}, however it is possible
to set up a full-fledged topological quantum computation using certain types of topological color codes in higher dimensional lattices \cite{self-correcting09}.
Under these conditions, it is possible to prove that self-correcting quantum computation including state preparation, quantum gates and measurement
can be carried out in the presence of the disturbing thermal noise. Additionally note that thermal noise does not always turn out detrimental in quantum information, even for systems without topological order \cite{WhiteNoise,SR}.

In this work we extend those results regarding the thermal effects on generalized toric codes constructed out of qudits.
We hereby summarize briefly some of our main results:

\noindent i/ We formulate the dynamics of a generalized toric code based on qudits at finite temperature.
To this end, we find the master equation coupling the qudits of the system code to a thermal bath.

\noindent ii/ We study and classify the different types of thermal processes that may occur when the anyonic excitations
are created, annihilated or diffused throughout the system. In particular, we find that for qutrits new types of anyons
and thermal processes appear that are forbidden for qubits.

\noindent iii/ The master equation is too involved so as to yield an explicit expression for the decay rate of the topological order initially present in the code.
However, in a short-time regime it is possible to solve it and find expressions for the decay rates as a function of the dimension $d$ of the qudits.
Interestingly enough, we find that the decay rate for qutrits presents a crossover temperature $T_c$ that is absent for any other qudits.

\noindent iv/ We can give an explicit proof that for times long enough, the non-local order parameter representing the topological order in the system decays to zero.

This paper is organized as follows: In Sect.\ref{sec:II} we review the formulation of the master equation of the 2D Kitaev code for qubits in order to establish the notation and
the necessary tools to study thermal effects in more general toric codes. We also introduce a non-local order parameter and study the fate of
topological orders for two different regimes: short time-regime and long-time regime.
In Sect.\ref{sec:III} we find the master equation describing topological qutrits coupled to a thermal bath. This allows us to see new energy processes for the
anyonic excitations that are not present when the toric code is made up of qubits. Likewise, the short-time regime has a different behavior that can be
seen in the initial decay rate of the topological order. In particular, we can define a crossover temperature for qutrits where the decay rate is better than
with other qudits.
Sect.\ref{sec:IV} is devoted to conclusions.
We refer to Appendix \ref{App-TOqutrits} for evolution of the order parameter for qutrits
and Appendix \ref{App-irreducibility} for a proof of the irreducibility of the computational representation of the $d$--Pauli group
needed to study the master equation in the long-time regime.

\section{Thermal Stability of the Kitaev 2-D Model}
\label{sec:II}

\begin{figure}[h]
\centering
\includegraphics[scale=0.2]{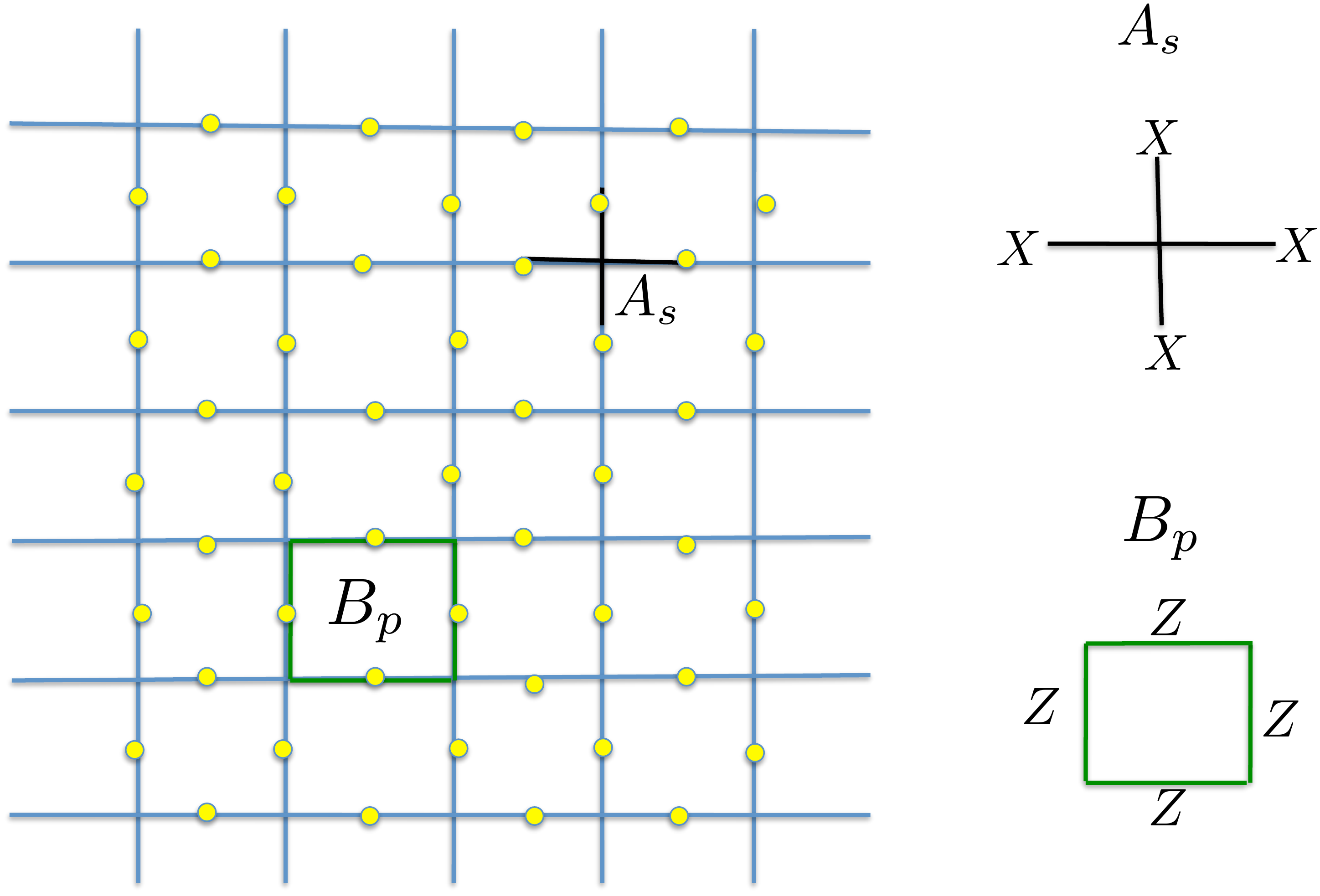}
\caption{Square lattice on the torus. The yellow points represent qubits.}
\label{Latticed2}
\end{figure}

We shall not dwell upon all details of  Kitaev's toric code~\cite{Kitaev}, however we will give the basic ideas to understand how to apply a thermal stability analysis to it,
as well as to establish the notation and methods.
We will consider a $k\times k$ square lattice embedded in a $2-$torus. Let us attach a qubit,  like a spin $1/2$, to each edge of the lattice. So we have $N=2k^2$ qubits. For each vertex $s$ and each
face $p$, we denote the stabilizer operators of the following form:
\begin{equation} \label{stabop}
  A_s\ :=\ \prod_{j\in\mbox{\scriptsize star}(s)}X_j,  \qquad\qquad
  B_p\ :=\ \prod_{j\in\mbox{\scriptsize boundary}(p)}Z_j,
\end{equation}
where $X_j$ and $Z_j$ are the Pauli Matrices applied to the qubit on site $j$. $A_s$ and $B_p$ commute among each other for they have either 0 or 2 common edges. They are also Hermitian and have eigenvalues $1$ and $-1$ (see figure~\ref{Latticed2}). Therefore, they constitute an abelian subgroup of the Pauli group of $n$ qubits that is a stabilizer group.

\begin{figure}[h]
\centering
\includegraphics[scale=0.3]{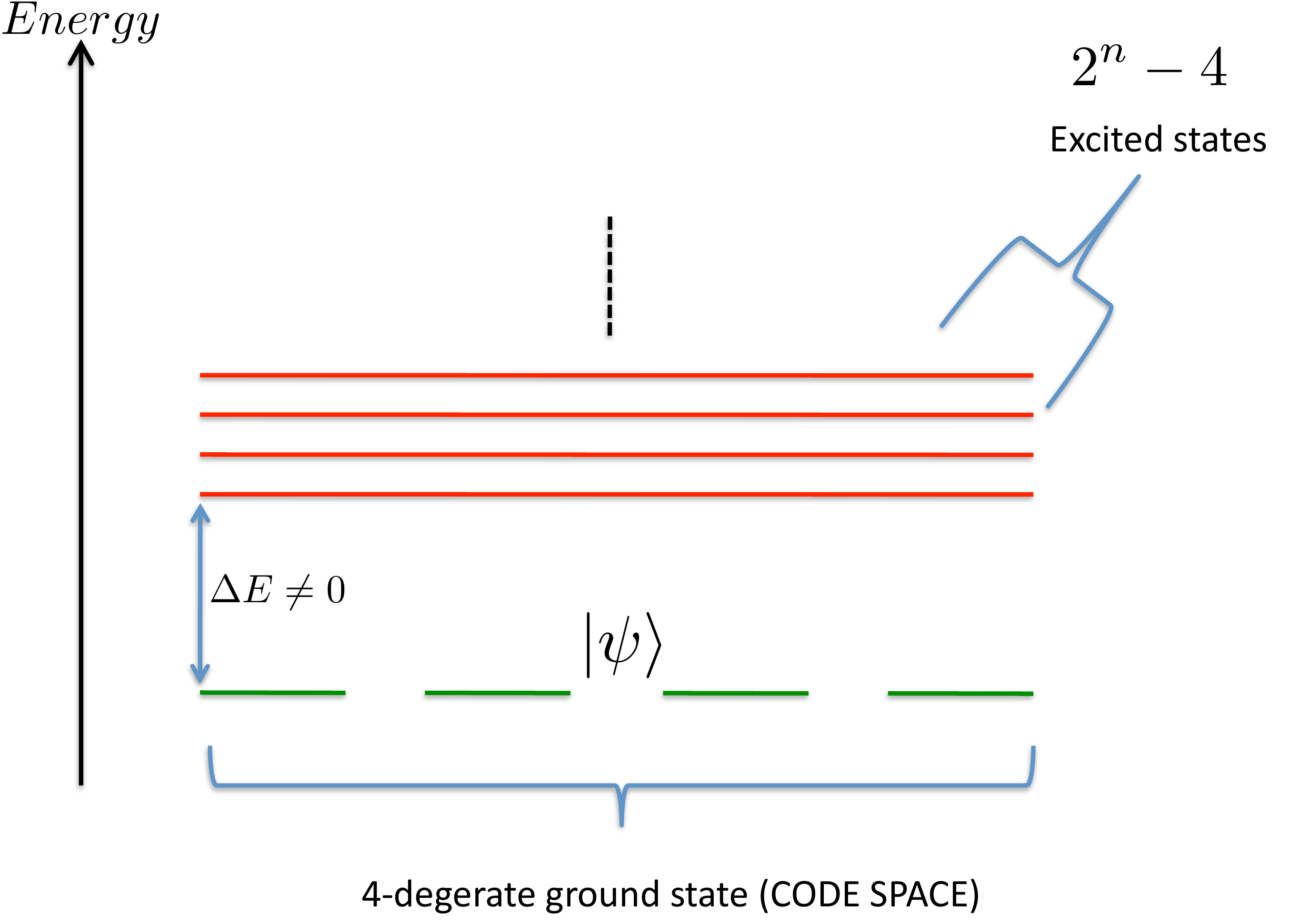}
\caption{Schematical spectrum of the Toric Code Hamiltonian. The ground state is the code space $\calC$ where we codify our information.}
\label{Energylev}
\end{figure}

Let $\cal H$ be the Hilbert space of all $n=2k^2$ qubits and define the topological quantum code or
\emph{protected subspace} $\calC\subseteq\calH$ as follows:
\begin{equation}
  \calC\ =\ \Bigl\{\, |\Psi\rangle\in\calH\,:\,\
  A_s|\Psi\rangle=|\Psi\rangle,\ B_p|\Psi\rangle=|\Psi\rangle\ \mbox{for all}\ s,p
  \,\Bigr\}.
\end{equation}\\
\indent This construction defines a quantum code called the \emph{toric code}.
The operators $A_s$, $B_p$ are the \emph{stabilizer operators} of this code, i.e. operators that leave trivially invariant the code space.
As we want to analyze the physical properties of this code, in particular the thermal properties of the topological order,
it is convenient  to define its associated Hamiltonian in the form:
\begin{equation}
  H^\mathrm{sys}\ :=\ -\sum_s A_s \,-\, \sum_p B_p.
  \label{H0}
\end{equation}
Complete diagonalization of  this Hamiltonian is possible since operators $A_s$, $B_p$ commute. In particular, the ground state coincides with the protected subspace of the code $\calC$; it is 4-fold degenerate (see figure~\ref{Energylev}).
All excited states are separated by an energy gap $\Delta E\ge 4$. This is due to the fact that the difference between the eigenvalues of $A_s$ $(B_p)$ is equal to 2.
Excitations come in pairs since they correspond to violations of the plaquette and/or vertex stabilizer operators and these must comply with the
overall constraints $\prod_s A_s=1$\, and $\prod_p B_p=1$.  Thus, excitations are represented as open strings in the direct or the dual lattice of the original square lattice.

An essential feature of this Hamiltonian is its locality in terms of four-body interactions, very useful for practical purposes.  Another key property is that this Hamiltonian model is gapped,
which led to the initial expectation that  all type of ``errors'', i.e.\ noise-induced excitations
will be removed automatically by some relaxation processes. Of course, this
requires cooling, i.e.  some coupling to a thermal bath with low temperature
(in addition to the Hamiltonian~(\ref{H0})) as we shall describe later on.
It can be shown that this Hamiltonian is robust under local quantum perturbations at zero temperature \cite{BT09}:  there would be a level splitting which will vanish as $\exp(-ak)$, where $k$ is the length of the lattice \cite{Kitaev}.

Due to this unavoidable coupling to a thermal bath, our system is subjected to thermal errors as well. These can be seen as violations on the plaquette and vertex conditions: $A_s|\Psi\rangle=|\Psi\rangle,\ B_p|\Psi\rangle=|\Psi\rangle $. Moreover, $A_s$ and $B_p$ are unitary, and also Hermitian in the case of qubits. Therefore, violations on the plaquette and/or vertex condition mean are given by\begin{equation}
A_s|\Psi\rangle=-|\Psi\rangle,~~~~~~~~~ B_p|\Psi\rangle=-|\Psi\rangle,
\end{equation}
for a certain number of  sites $s$ and/or plaquettes $p$.

These violations cost energy to our system, thereby becoming excitations. And as long as they always come in pairs  (to satisfy the condition $\prod_s A_s=1$\, and $\prod_p B_p=1$), they can be seen (pictorially) as string operators with plaquette or vertex violations at the ends.

Errors on the system  can be expressed in terms of operators $\sigma^x$, $\sigma^z$, or products among themselves. These operators act on each edge $j$ where the physical qubits are placed. We use the notation $\sigma^x$ for a Pauli operator of type $X$ when it is referred to an error, i.e., a bump operator acting due to the coupling to the thermal bath. Similarly with  $\sigma^z$. Namely,  it is just a matter of notation to distinguish when we have an operator that defines our stabilizer operators in $A_s, B_p$ and when we have an error acting on the system. To see what is the effect that they produce, we will see how the ground state changes by applying these $\sigma^{x,z}$. We will see that this corresponds to creation, annihilation  and movement of a pair of excitations, that from now on we shall refer to them as \emph{anyons}. They are called anyons since their wave function picks up a different phase than fermions or bosons when we exchange the end-particles of string operators of $x$-type with $z$-type.
According to this notation, when we apply a bump operator from the thermal bath,  it will act on the ground state of the system as follows:
\begin{equation}
\sigma_j^z\ket{\Psi},
\end{equation}
where $\ket{\Psi}$ is the ground state of the system where our information is encoded. This means that the physical qubit at the edge $j$ has been bumped.
The energy cost will be $\Delta E=4$ in energy units of the system corresponding to  the definition of $H^\mathrm{sys}$ .

As a first step, one is interested in designing a stable quantum memory, i.e.\ a $N-$particle system which can support at least a single encoded logical qubit for a long time, preferably with this time growing exponentially with $N$. This is the notion of stability  that  we shall refer to from now on. In the paper \cite{Horodecki} by Alicki \emph{et al.} they provide a rigorous method to prove thermal instability of the 2D Kitaev model and obtain a master equation that describes the dynamics of the system weakly coupled to a thermal environment. We will study the problem of thermal instability within the framework of topological orders obtaining complementary and interesting results.

\subsection{Davies' Formalism}

Let us consider a small and finite system, that is coupled to one or more heat baths at the same inverse temperature $\beta=(k_{\rm B}T)^{-1}$ leading to the total Hamiltonian
\begin{equation}
H = H^\mathrm{sys} + H^{\text{bath}} + V \enskip\text{with}\enskip V = \sum_\alpha S_\alpha \otimes f_\alpha.
\end{equation}
Here $H^\mathrm{sys}$ represents the Hamiltonian of the system, where the quantum information is encoded and which we want to protect from the external thermal noise. $H^{\text{bath}}$ is the bath Hamiltonian, i.e., it describes the internal dynamics of the bath which is out of our control. Finally $V$ represents the coupling between the system and the thermal bath. $S_\alpha$ and $f_\alpha$ are operators which act on the system and bath respectively. Both the coupling operators $S_\alpha$ and $f_\alpha$ are assumed to be Hermitian (without loss of generality \cite{Libro}).

In the weak coupling regime that we shall assume throughout this work, the Fourier transform $\hat g_\alpha$ of the auto-correlation function of $f_\alpha$ plays an important role as it describes the rate at which the coupling is able to transfer energy between the bath and the system \cite{Davies,AlickiLendi,Libro,Breuer}. Often a minimal coupling to the bath is chosen, minimal in the sense that the interaction part of the Hamiltonian is as simple as possible but still addresses all energy levels of the system Hamiltonian in order to have an ergodic reduced dynamics. This last condition is ensured if \cite{Frigerio,Spohn1,Spohn2,SpohnReview,Libro}
\begin{equation}\label{ergodicity}
\bigl\{ S_\alpha, H^\mathrm{sys} \bigr\}' = \mathds{C} \mathds{1},
\end{equation}
i.e.\ no system operator apart from those proportional to the identity commutes with all the $S_\alpha$ and $H^\mathrm{sys}$.

The weak coupling limit results \cite{Davies,AlickiLendi,Libro,Breuer} in a Markovian evolution for the system given in Heisenberg picture by the master equation
\begin{equation}
\frac{dX}{dt} = \c G(X) := {\rm i}\delta(X) + \c L(X).
\label{reddyn2}
\end{equation}
The generator of the evolution $\c G(X)$ is a sum of two terms, the first is a usual Liouville-von~Neumann term as in the quantum mechanics of closed systems, while the second is a particular type of Kossakowski-Lindblad generator:
\begin{align}
\delta(X)
&= [H^\mathrm{sys},X] \\
\c L(X)
&= \sum_\alpha \sum_{\omega\ge0} \c L_{\alpha\,\omega}(X)
\label{dis} \\
&= \sum_\alpha \sum_{\omega\ge0} \hat g_\alpha(\omega)\, \Bigl\{ \bigl(S_\alpha(\omega)\bigr)^\dagger\, \bigl[X \,,\, S_\alpha(\omega) \bigr] + \bigl[ \bigl( S_\alpha(\omega) \bigr)^\dagger \,,\, X \bigr]\,  S_\alpha(\omega)
\nonumber \\
&\phantom{=\ } + \r e^{-\beta\omega}\, S_\alpha(\omega)\, \bigl[ X \,,\, \bigl( S_\alpha(\omega) \bigr)^\dagger \bigr] + \r e^{-\beta\omega}\, \bigl[ S_\alpha(\omega) \,,\, X \bigr]\, \bigl( S_\alpha(\omega) \bigr)^\dagger \Bigr\}.
\label{davies}
\end{align}
Here the $S_\alpha(\omega)$ are the Fourier components of $S_\alpha$ as it evolves under the system Hamiltonian
\begin{equation}
\r e^{{\rm i}tH^\mathrm{sys}}\, S_\alpha\, \r e^{-{\rm i}tH^{\mathrm{sys}}} = \sum_\omega S_\alpha(\omega)\, \r e^{-{\rm i}\omega t},
\end{equation}
where the $\omega$'s are Bohr frequencies of the system Hamiltonian ($\hbar\omega=E_1-E_2$, for two energy levels $E_1$ and $E_2$).

In addition, the temperature of the environment appears in~(\ref{davies}) through $\beta$, and this generator is the so-called Davies generator \cite{Davies}, or  Born-Markov generator in the quantum optics literature.

\subsection{Master Equation for 2-D Kitaev Model with Qubits}

\begin{figure}[h]
\centering
\includegraphics[scale=0.4]{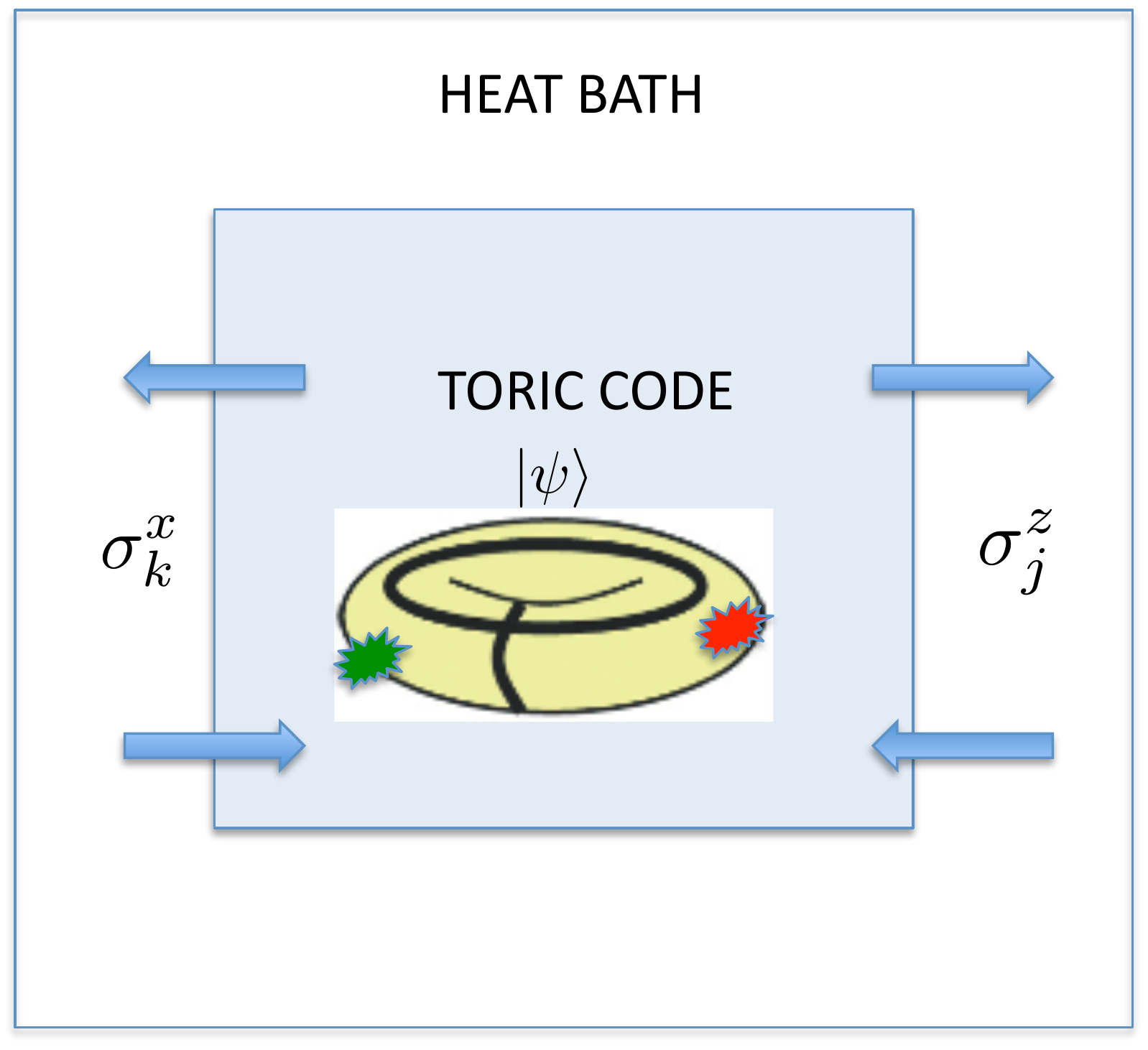}
\caption{Toric Code coupled to a heat bath. Outgoing arrows in the upper part of the figure mean information flowing from system-to-bath, and ingoing arrows in the lower part mean information flowing from bath-to-system.}
\label{heatbath}
\end{figure}

Given the simplicity of the Kitaev's model, we can apply the Davies' theory for studying its stability in the presence of thermal noise. This is pictorially represented in figure~\ref{heatbath}.

\begin{figure}[h]
\centering
\includegraphics[scale=0.3]{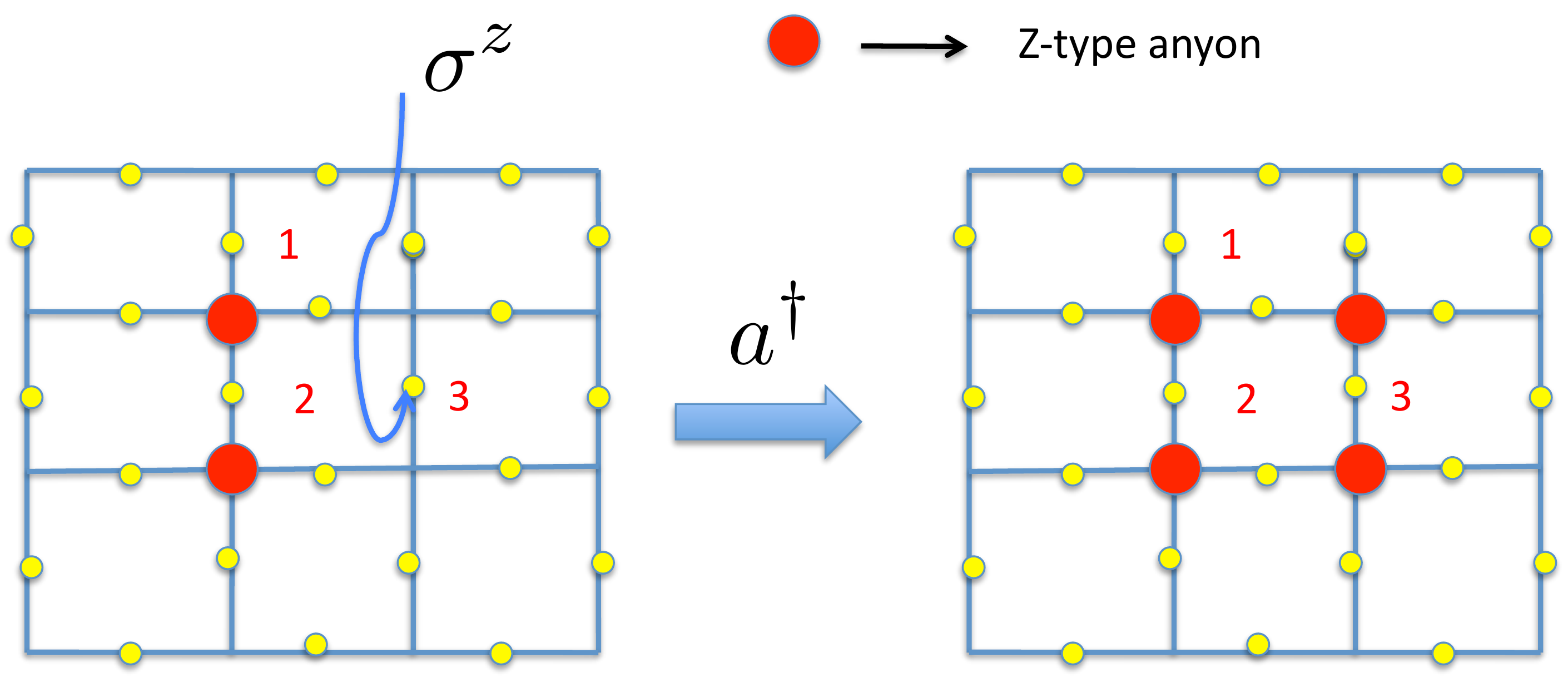}
\caption{Dynamics induced by the heat bath on the Toric Code. Creation of a new pair of anyons. Energy increases $\Delta E=4$.}
\label{any3}
\end{figure}

\begin{figure}[h]
\centering
\includegraphics[scale=0.3]{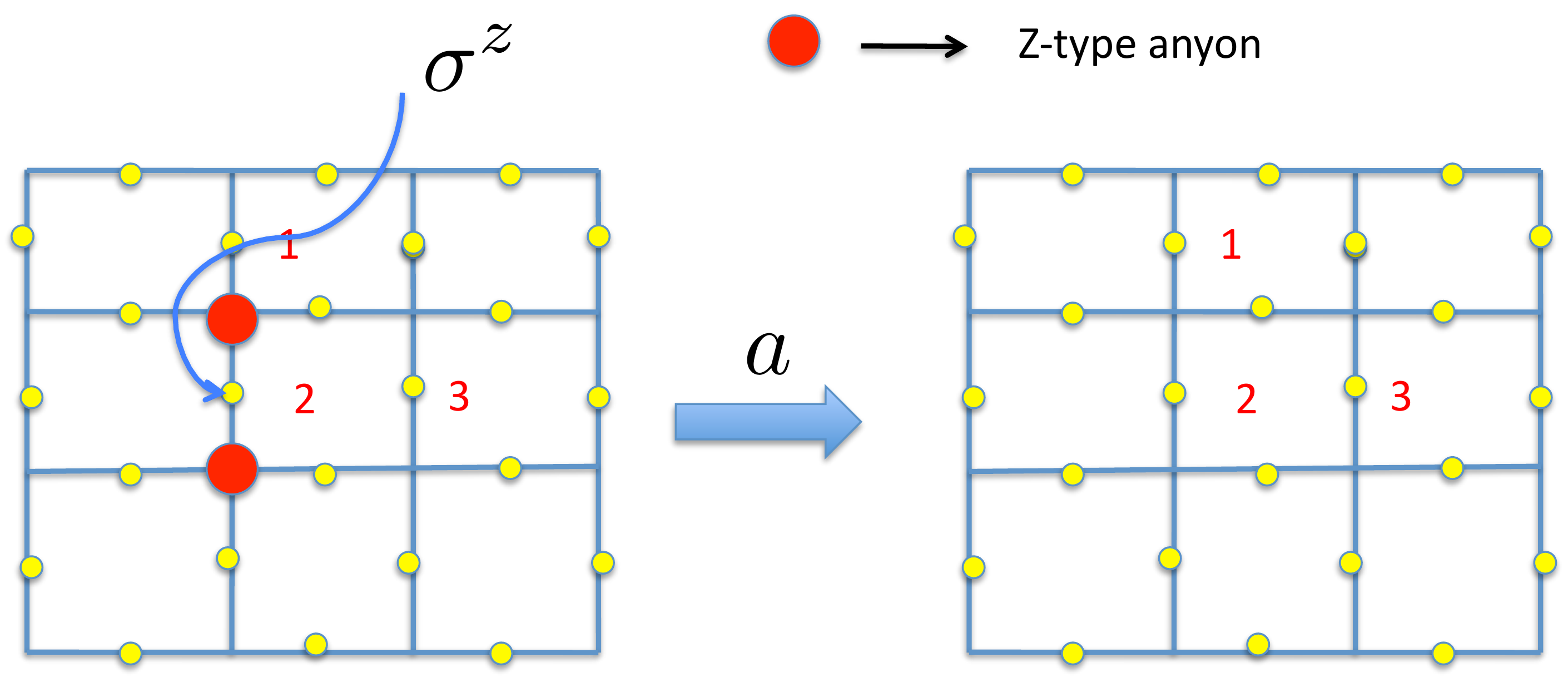}
\caption{Dynamics induced by the heat bath on the Toric Code. Annihilation of a pair of anyons. Energy goes down by $\Delta E=4$.}
\label{any2}
\end{figure}

\begin{figure}[h]
\centering
\includegraphics[scale=0.3]{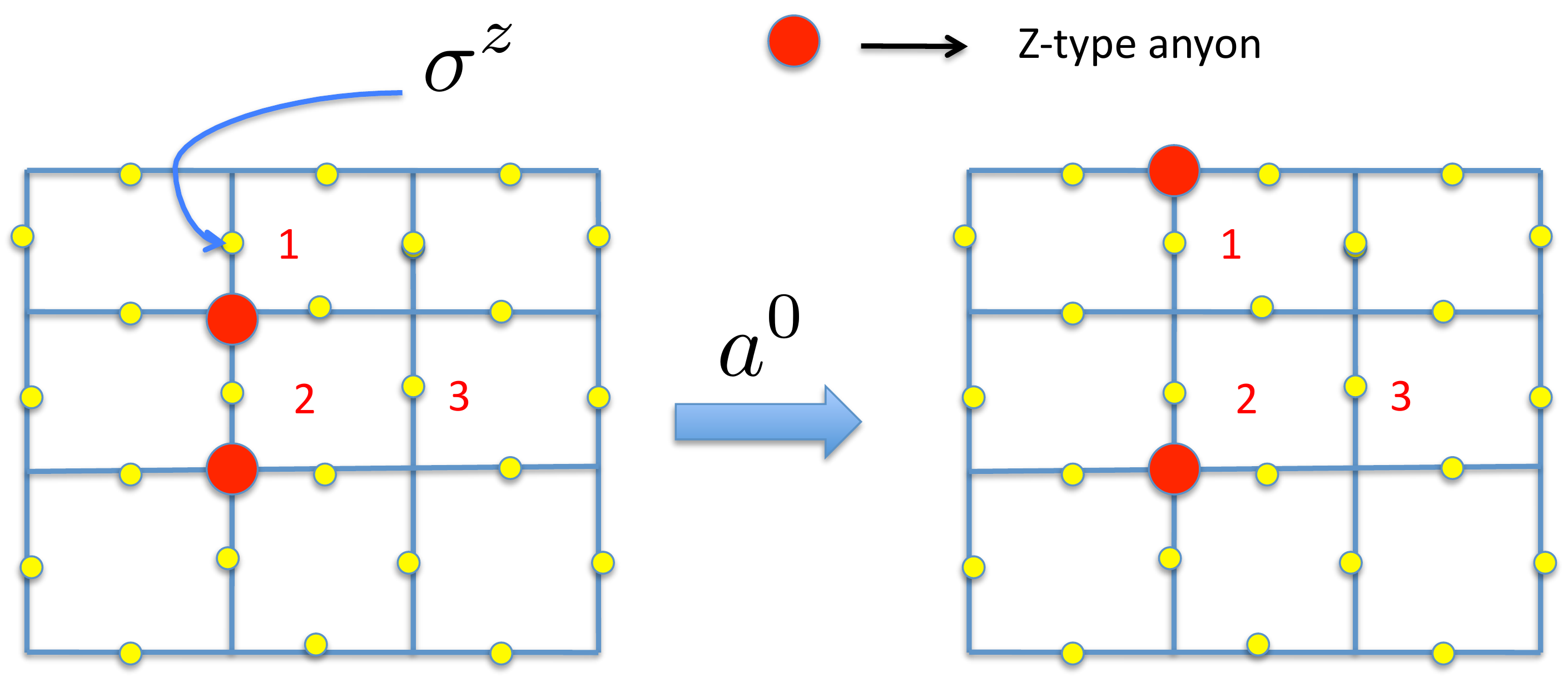}
\caption{Dynamics induced by the heat bath on the Toric Code. Pure decoherence by moving an anyon with  no energy change.}
\label{any1}
\end{figure}

The interaction Hamiltonian is assumed to be local and associated with $\sigma^x$ and $\sigma^z$ errors:
\begin{equation}\label{V}
V=\sum_j\sigma^x_j\otimes f_j^x+\sigma^z_j\otimes f_j^z,
\end{equation}
where $f_j^x$ and $f_j^z$ are associated with two different baths. Thus, first of all, we need to compute the Fourier transform of the system operators $\r e^{{\rm i}tH^{\mathrm{sys}}}\sigma^x_j\r e^{-{\rm i}tH^\mathrm{sys}}$ and $\r e^{{\rm i}tH^{\mathrm{sys}}}\, \sigma^z_j\, \r e^{-{\rm i}tH^\mathrm{sys}}$ in order to define the dynamical operators of the system. Here  $H^\mathrm{sys}:=H_0=\ -\sum_s A_s \,-\, \sum_p B_p$, with $[A_s,B_p]=0$, $[A_s,\sigma^z_j]=0$ and $[B_p,\sigma^x_j]=0$. Thus,  stabilizers $A_s$ only play a role in the Fourier transform of $\sigma^x_j$ and $B_p$ only in $\sigma^z_j$. By computing this Fourier transform, we obtain the dynamical operators of the system due to the coupling to the thermal bath. With $\Delta=4$ denoting the gap of the Toric Code Hamiltonian,  then the expression of these operators $S_{\alpha}(\omega)$  that appear in equation~\eqref{davies} are \cite{Horodecki} :

\begin{enumerate}
\item \emph{Operators associated with $\sigma^x_j$ errors}:
\begin{eqnarray}\label{ErrorOp2x}
S^x_j(0)&:=&b_j^0=\sigma^x_jR_j^0, \nonumber \\
S^x_j(\Delta)&:=&b_j=\sigma^x_jR_j^{+}, \\
S^x_j(-\Delta)&:=& b_j^{\dagger}=\sigma^x_jR_j^{-}, \nonumber
\end{eqnarray}
with  $R_j^0:=\frac{1}{2}(1-B_pB_{p'})$ and $R_j^{\pm}:=\frac{1}{4}(1\mp B_p)(1\mp B_{p'})$ being orthogonal projectors.
\item \emph{Operators associated with $\sigma^z_j$ errors}:
\begin{eqnarray}\label{ErrorOp2z}
S^z_j(0)&:=&a_j^0=\sigma^z_jP_j^0,\nonumber\\
S^z_j(\Delta)&:=&a_j=\sigma^z_jP_j^{+},\\
S^z_j(-\Delta)&:=&a_j^{\dagger}=\sigma^z_jP_j^{-},
\end{eqnarray}
and the projectors: $P_j^0:=\frac{1}{2}(1-A_sA_{s'})$ and $P_j^{\pm}:=\frac{1}{4}(1\mp A_s)(1\mp A_{s'})$.
\end{enumerate}
\vspace{0.5cm}
\indent These operators have a nice interpretation in terms of anyonic properties of the system:
\begin{enumerate}
\item $a_j^{\dagger}(b_j^{\dagger})$ creates a pair of anyons of $z$-type($x$-type) on the lattice at position $j$. See figure~ \ref{any3}.
\item $a_j(b_j)$ annihilates a pair of anyons of $z$-type($x$-type) on the lattice at position $j$. See figure~\ref{any2}.
\item $a_j^0(b_j^0)$ moves a pair of anyons of $z$-type($x$-type) on the lattice. See figure~\ref{any1}.
\end{enumerate}

\smallskip

Thus, the dissipator of the master equation $\c L(X)$ for the system is:
\begin{eqnarray}
\mathcal{L}(X)&=&\mathcal{L}^z(X)+\mathcal{L}^x(X),\nonumber\\
\mathcal{L}^x(X)&=&\sum_j\frac{1}{2}R(4)\{(-b^{\dagger}_jb_jX-Xb^{\dagger}_jb_j+b^{\dagger}_jXb_j)+\text{e}^{-4\beta}(-b_jb^{\dagger}_jX-Xb_jb^{\dagger}_j+\nonumber\\
&+&b_jXb^{\dagger}_j\}-\frac{1}{2}R(0)[b^0_j,[b^0_j,X]],\nonumber\\
\mathcal{L}^z(X)&=&\sum_j\frac{1}{2}R(4)\{(-a^{\dagger}_ja_jX-Xa^{\dagger}_ja_j+a^{\dagger}_jXa_j)+\text{e}^{-4\beta}(-a_ja^{\dagger}_jX-Xa_ja^{\dagger}_j+\nonumber\\
&+&a_jXa^{\dagger}_j\}-\frac{1}{2}R(0)[a^0_j,[a^0_j,X]].\nonumber
\end{eqnarray}
Where $R(4)$ and $R(0)$ are the exchange rate between the system and the bath associated to each Bohr frequency, namely $\omega=0,4$, assuming units of $J=1$.\\

\subsection{Topological Order}

We shall study the evolution of the expectation value $\bra{{\rm GS}}X_c\ket{{\rm GS}}$ as a simple order parameter, where $X_c$ is the tensor product of $\sigma^x$ Pauli operators along one non-contractible loop on the surface of the torus and $\ket{{\rm GS}}$ denotes a generic ground state of the system Hamiltonian. This ground state is a superposition of the degenerate states in the ground state manifold of $H^{{\rm sys}}$, namely ${\calC}$. This gives us a sufficient measure of the topological order of the system \cite{Nussinov}. If this quantity falls to zero during the time evolution for every element of ${\calC}$, there is not a global and self-protected way to encode quantum information.
The evolution of the operator $X_c$ is given by equation \eqref{reddyn},
\[
\frac{dX_c(t)}{dt}={\rm i}\delta[X_c(t)]+\mathcal{L}[X_c(t)].
\]
In order to simplify the computation, we remove the free evolution by performing the transformation
\begin{equation}
\tilde{X}_c(t)=\r e^{-{\rm i}H^{\mathrm{sys}}t}X_c(t)\r e^{{\rm i}H^\mathrm{sys}t}.
\end{equation}
Since the dissipator is invariant under this transformation, we obtain
\begin{equation}\label{master}
\frac{d\tilde{X}_c(t)}{dt}=\mathcal{L}[\tilde{X}_c(t)].
\end{equation}
Interestingly, for the expectation value we obtain $\bra{{\rm GS}}X_c(t)\ket{{\rm GS}}=\bra{{\rm GS}}\tilde{X}_c(t)\ket{{\rm GS}}$, as $\ket{{\rm GS}}$ is an eigenstate of $H^{\mathrm{sys}}$.

\noindent Taking into account expressions (\ref{ErrorOp2x}) and (\ref{ErrorOp2z}) , the action of the dissipators on $X_c$ can be simplified to
\begin{equation}\label{Lx}
\mathcal{L}_x(X_c)=-\frac{1}{2}\sum_jR(\Delta)\left([R^j_+,[R^j_+,X_c]]+\r e^{-\Delta\beta}[R^j_-,[R^j_-,X_c]]\right)+R(0)[R^j_0,[R^j_0,X_c]],
\end{equation}
and
\begin{eqnarray*}
\mathcal{L}_z(X_c)&=&\sum_jR(\Delta)[P_+^j \sigma^z_j X_c \sigma^z_j P_+^j-P_+^jX_c+\r e^{-\Delta\beta}(P_-^j \sigma^z_j X_c \sigma^z_j P_-^j-P_-^jX_c)]\\
&+&R(0)[P_0^j \sigma^z_j X_c \sigma^z_j P_0^j-P_0^jX_c],
\end{eqnarray*}
where we have used the fact that $[P_{\pm,0}^j,X_c]=0$ for every $j$, as these projectors are only functions of vertex operators. However, the same assertion is not true for $R_{\pm,0}^j$ in general. If $j\not\in c$, i.e. $j$ does not belong to the path where $X_c$ is acting on, every element commutes with each other and their contribution is zero. On the other hand, if $j\in c$, as $\sigma^z_j\sigma^x_j\sigma^z_j=-\sigma^x_j$,  the string operator yields $\sigma^z_jX_c\sigma^z_j=-X_c$. Therefore,  simplifying we obtain
\begin{equation}\label{Lz}
\mathcal{L}_z(X_c)=-\frac{\Delta}{2}|c|X_c\{R(\Delta)[P_+^j+\r e^{-\Delta\beta}P_-^j]+R(0)P_0^j\},
\end{equation}
where $|c|$ is the number of points in the path $c$.

\subsection{Short-Time Regime}

The solution to the master equation (\ref{master}) is formally written as $\tilde{X}_c(t)=\r e^{\mathcal{L}t}X_c$. However, this expression is too involved to be computed analytically except for short and long times to be specified hereby. In the first case, at lowest order we have
\begin{equation}
\tilde{X}_c(t)\simeq(1+t\mathcal{L})X_c.
\end{equation}
The evolution of $\bra{{\rm GS}}X_c(t)\ket{{\rm GS}}$ is given by
\begin{equation}
\langle\tilde{X}_c(t)\rangle\simeq[1-2t|c|R(\Delta)\r e^{-\Delta\beta}]\langle X_c(0)\rangle.
\end{equation}
To arrive at this equation, we have used the fact that for all $j$:
\begin{eqnarray*}
P_{+,0}^j\ket{{\rm GS}}&=&0,\\
P_{-}^j\ket{{\rm GS}}&=&\ket{{\rm GS}},\\
R_{+,0}^j\ket{{\rm GS}}&=&0,\\
R_{-}^j\ket{{\rm GS}}&=&\ket{{\rm GS}}.
\end{eqnarray*}
Thus, the contribution of $\mathcal{L}_x$ is zero:
\begin{eqnarray}
\bra{{\rm GS}}\mathcal{L}_x(X_c)\ket{{\rm GS}}&=&-\frac{1}{2}\sum_jR(\Delta)\r e^{-\Delta\beta}\bra{{\rm GS}}[R^j_-,[R^j_-,X_c]]\ket{{\rm GS}}\nonumber\\
&=&-\frac{1}{2}\sum_jR(\Delta)\r e^{-\Delta\beta}\left(\bra{{\rm GS}}[R^j_-,X_c]\ket{{\rm GS}}-\bra{{\rm GS}}[R^j_-,X_c]\ket{{\rm GS}}\right)=0.
\end{eqnarray}
Whereas for $\mathcal{L}_z$, we have
\begin{equation}
\mathcal{L}_z(X_c)=-\frac{\Delta}{2}|c|R(\Delta)\r e^{-\Delta\beta}\bra{{\rm GS}}X_c\ket{{\rm GS}}.
\end{equation}

\noindent Finally, as $\bra{{\rm GS}}X_c(t)\ket{{\rm GS}}=\bra{{\rm GS}}\tilde{X}_c(t)\ket{{\rm GS}}$, the desired equation valid at short times is
\begin{equation}\label{Gammaqubit}
\langle X_c(t)\rangle\simeq\left[1-\frac{\Delta}{2}t|c|R(\Delta)\r e^{-\Delta\beta}\right]\langle X_c(0)\rangle,
\end{equation}
with $\Delta=4$.

It is important to remark that $R(0)$ does not appear in the initial decay rate, as long as  short times is concerned. The diffusion of anyons is a second-order process in time as it requires first the creation of a pair of anyons with $R(\Delta)$, and later the free diffusion with $R(0)$.

\subsection{Long-Time Regime}

On the other hand, in order to analyse the thermal properties for long times, we write the Davies generator in the Schr\"odinger picture through the relation $\mathrm{Tr}[\mathcal{L}^\dagger(\rho)X]=\mathrm{Tr}[\rho\mathcal{L}(X)]$ for any $X$ and $\rho$. It is a well-known result \cite{Davies,AlickiLendi,Libro,Breuer} that the Gibbs state is a stationary state for $\mathcal{L}^\dagger$,
\begin{equation}
\mathcal{L}^\dagger(\rho_\beta)=0,
\end{equation}
where $\rho_\beta=\r e^{-\beta H^\mathrm{sys}}/Z$,  $\beta$ is the same to the inverse temperature as the surrounding bath, and $Z:=\mathrm{Tr}(\r e^{-\beta H^\mathrm{sys}})$ is the system partition function . To guarantee  that any initial state of the system relaxes to $\rho_\beta$, we can resort to  condition (\ref{ergodicity}).
In our case this follows from the Schur's lemma as $S_\alpha=\sigma^x_j,\sigma^z_j$ and $\{\mathds{1},\sigma^x,\sigma^z,\sigma^x\sigma^z\}$ form an irreducible representation of the Pauli group.

Thus $\bra{{\rm GS}}X_c(t)\ket{{\rm GS}}\simeq\mathrm{Tr}[X_c\rho_\beta]$ for large $t$, and we have $\mathrm{Tr}[X_c\rho_\beta]=0$. This is simply due to the fact that $\rho_\beta$ is diagonal in any of the possible eigenbasis of $H^\mathrm{sys}$, and it is not difficult to choose one such that $X_c$ vanish on diagonal elements,
\begin{equation}
\mathrm{Tr}[X_c\rho_\beta]=\frac{1}{Z}\sum_i\r e^{-\beta\lambda_i}\bra{\psi_i}X_c\ket{\psi_i}=0,
\end{equation}
for some eigenbasis $\{\ket{\psi_i}\}$ of $H_0$, the Kitaev's Hamiltonian.

In conclusion whatever the initial value of the order parameter $\langle X_c(0)\rangle$ is, it decays to zero during the time evolution of the system provided that the temperature is finite. The decay rate at short times is equal to $\frac{\Delta}{2}|c|R(\Delta)\r e^{-\Delta\beta}$. Note the detrimental effect of the factor $|c|$: the larger size of the system, the larger the decay rate. In order to keep the order parameter above certain finite value such that $\langle X_c(0)\rangle\neq0$, this decay rate must decrease, which is not the case when increasing the system size.

\section{Kitaev 2-D Model for Qudits}
\label{sec:III}

In this section we consider again a 2-D toric code, but instead of assuming that we have a two-level system  on each site, we will consider that particles arranged on the torus, have $d$ accessible levels. We will first derive a general theory for \emph{qudits} and then consider the case $d=3$ (qutrits). A qutrit can be represented for instance, as a particle of spin 1 or  a 3-level system in an atom, etc.

This problem is very  interesting since qutrits have certain advantages with respect to qubits, namely:
\begin{enumerate}
\item Larger capacity of information storage.
\item Quantum channels are more robust for qutrits. For example: Bell inequalities are proved with more accurate bounds. This is relevant for Quantum Key distribution.
\item Entanglement quantum destillation is more efficient with qutrits than with qubits \cite{Bombin}.
\item Qutrit logic gates \cite{Yao} are also capable of providing universal quantum computation. i.e.,  the necessary computational power to construct all possible logic gates~\cite{Gottesman}.
\end{enumerate}

To build a system like that, we will try to choose the Hamiltonian and the operators acting on the system in the same way as before. Previously, for two-level systems, we have considered the Pauli matrix algebra to be the basis of operators in our system.  Now, we have to use
a proper generalization  for dimension $d$. As $iXZ=Y$  gives the second Pauli matrix, it is enough to consider $X$ and $Z$ in this generalization to quantum states with $d$ multilevels. However,  the generalization of  Pauli matrices to dimension $d$ is not unique \cite{notunique}. Thus, we shall select the most important properties of Pauli matrices of dimension 2 for our purpose of quantum error correction.

In $d=2$ we defined a basis: $\ket{0},\ket{1}$ in the Fock space of each particle. They are defined as the eingenstates of the $Z$ Pauli matrix. And the $X$ Pauli matrix takes $\ket{0}$ to $\ket{1}$ and viceversa.
\begin{equation*}
X=\left( \begin{array}{lll} 0 & 1 \\ 1 & 0  \end{array}\right),~~ Z=\left( \begin{array}{lll} 1 & 0 \\ 0 & -1  \end{array}\right),
\end{equation*}
\begin{eqnarray}
Z\ket{0}&=&+\ket{0},\qquad Z\ket{1}=-\ket{1},\nonumber\\
X\ket{0}&=&+\ket{1},\qquad X\ket{1}=\ket{0},\nonumber\\
\nonumber
\end{eqnarray}

The key important properties of these matrices for doing error correction are:
\begin{itemize}
\item They satisfy a cyclic condition (i.e. applying twice $Z$ or $X$ Pauli matrices is the identity) i.e. they are unitary.
\item They anticommute, that means $XZ=-ZX$.
\end{itemize}

Those are the properties that are generalized  to the $d-$dimensional case. Hermiticity is not taken into account as a basic ingredient, as we can always add the Hermitian conjugate obtaining an Hermitian operator, e.g. $\tilde{Z}=Z+Z^{\dagger}$, then $\tilde{Z}$ is Hermitian. Now we consider a basis for the particle Fock space: $\ket{0},\ket{1},...,\ket{d-1}$ which will be the eingenvectors of the  generalized $Z$ matrix with a certain eingenvalue. We define $X$ as the operator which takes the state $\ket{0}$ to $\ket{1}$ then $\ket{1}$ to $\ket{2}$ and so on. We will also ask for a cyclic condition as in the previous case:
\begin{equation}
\displaystyle X^d=\mathds{1}, \quad Z^d=\mathds{1}.
\label{cyclic}
\end{equation}
All these requirements can be cast on to the following defining relations:
\begin{eqnarray}
Z\ket{0}&=&+\ket{0},~~~~Z\ket{1}=\omega\ket{1}~~~~Z\ket{2}=\omega^2\ket{2},~~~~...~~~~,Z\ket{d-1}=\omega^{d-1}\ket{d-1};\nonumber\\
X\ket{0}&=&+\ket{1},~~~~X\ket{1}=\ket{2},~~~~...~~~~,X\ket{d-1}=\ket{0}.\nonumber\\
\label{opd3}
\end{eqnarray}

\noindent Looking at equation~(\ref{opd3}) we can deduce the meaning of operators $X$ and $Z$. $X$ is the displacement operator in the computational basis (i.e. in the Fock space basis of the physical qudits). $Z$ is the dual operator of $X$ under a discrete Fourier transform. In other words, $Z$ is diagonal in the computational basis and its eingenvalues are the weights of the Fourier transform. Thus, $X$ plays the role of the displacement operator and $Z$ is the dual operator on a system with discrete states of qudits \cite{Gottesman}.

\noindent Due to the cyclic condition (\ref{cyclic}) of $Z$ ($Z^d=\mathds{1}$) we have the relation $\omega^d=1$ where in general $\omega$ is a complex number. This implies that $\omega$ is a primitive $d-$root of the unity,
\begin{equation} \omega=\r e^{{\rm i}\frac{2\pi}{d}}. \end{equation}
Additionally, we can easily verify that $ZX=\omega XZ$, as follows from equation~(\ref{opd3}).

We  have already the algebra of operators that we are going to use  in order to built the stabilizer operators on this qudit toric code. The problem is that if we construct the vertex and plaquette operators as before, namely,
\begin{equation} \label{stabop3}
  A_s\ =\ \prod_{j\in\mbox{\scriptsize star}(s)}X_j,  \qquad\qquad
  B_p\ =\ \prod_{j\in\mbox{\scriptsize boundary}(p)}Z_j,
\end{equation} \\
then  $[A_s,B_p]\not=0$ for all \emph{s} and \emph{p}. They commute with each other provided that they do not share any common edge, but that is not the case if they share two. This happens because in this case the operators $X$ and $Z$ are no longer Hermitian.

\begin{figure}[h]
\centering
\includegraphics[scale=0.3]{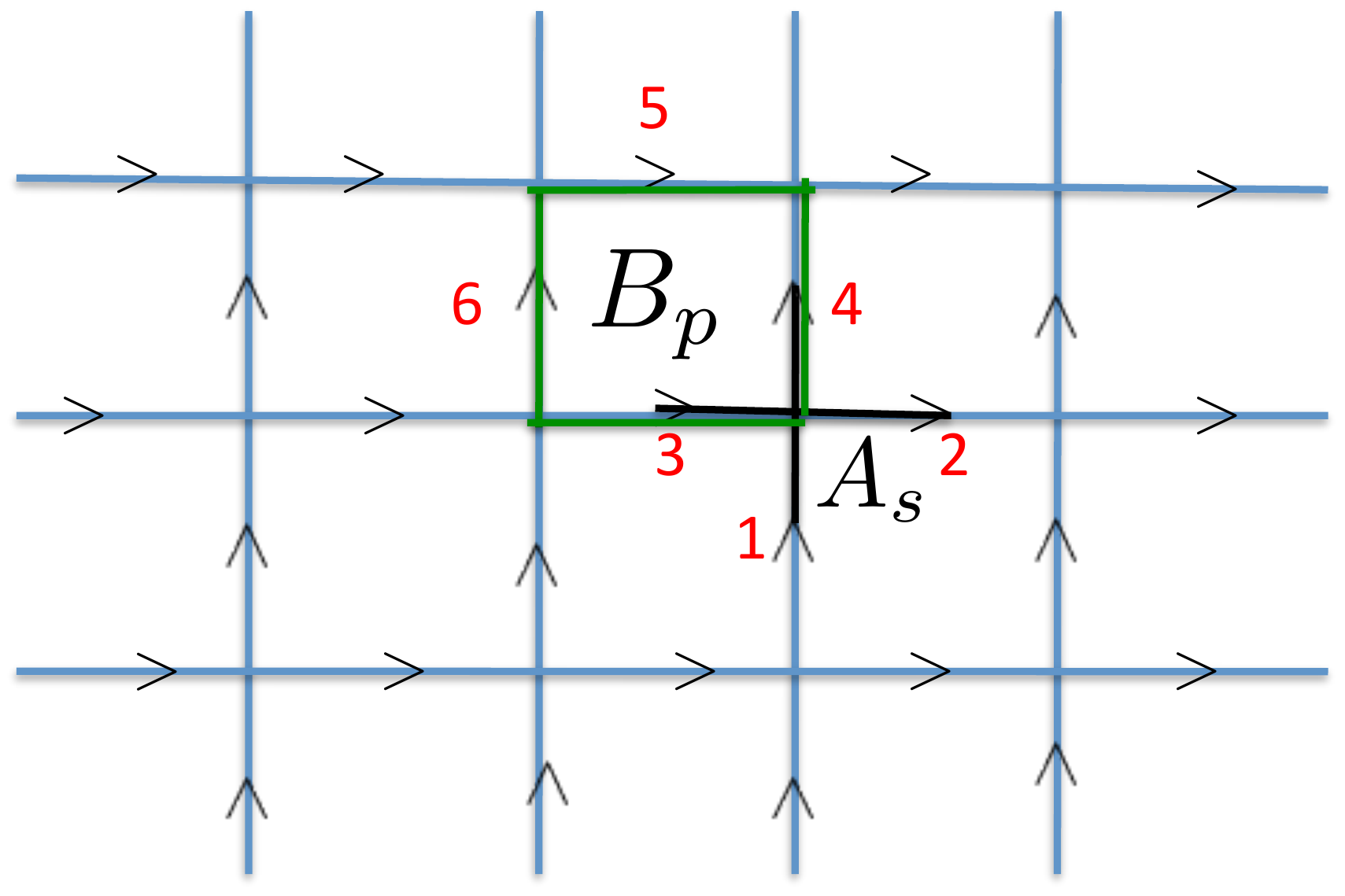}
\caption{New lattice for qudits showing vertex $A_s$ and plaquette $B_p$ operators: orientation of the lattice is necessary.}
\label{Operatorsd3}
\end{figure}

As shown in figure~\ref{Operatorsd3}, we have
\begin{equation}
[A_s,B_p]=[X_1X_2X_3X_4,Z_3Z_4Z_5Z_6]=(1-\omega^2)A_sB_p,
\end{equation}
which does no vanish for for dimension $d>2$, $(1-\omega^2)\not=0$. The case of $d=2$ is a very special case with $\omega=-1$ and therefore $(1-\omega^2)=0$. This happens because for $d=2$, $X$ and $Z$ are Hermitian operators. We need to think of another way to define our operators to have the same commutation rules as before,
and this leads to define an orientation on the lattice. This is shown in figure~\ref{Operatorsd3}. Defining an orientation on the lattice is a direct consequence of the non-Hermiticity of operators $X$ and $Z$.

Using the orientation of the lattice, we define the stabilizer operators in the following way. To build the vertex operators $A_s$ we assign an operator $X$ or $X^{-1}$ depending on the arrows of the edges of the lattice. If an arrow is pointing towards the vertex $j$, we will use $X^{-1}_j$ to build $A_s$, and if the arrow is pointing out another vertex we use $X_j$. For plaquette operators $B_p$, $Z_k$ is taken if the arrow is pointing clockwise and $Z^{-1}_k$ for anti-clockwise as shown in figure ~\ref{Operatorsd3}.
To see now that we obtain the correct commutation rule, we look again at figure~\ref{Operatorsd3} and check,
\begin{equation}
[A_s,B_p]=[X_1^{-1}X_2X_3^{-1}X_4,Z_3^{-1}Z_6Z_5Z_3^{-1}]=(1-\omega\omega^{-1})A_sB_p=0
\end{equation}
Then, the Hamiltonian could be written as follows:
\begin{equation} \label{Ham}
  H_\mathrm{aux}\ :=\ -\sum_s A_s \,-\, \sum_p B_p.
\end{equation}
Although, according to the definition of $A_s$ and $B_p$, this operator is unitary, it is important to note that the operators $A_s$ and $B_p$ are not Hermitian any more, so $H_\mathrm{aux}$ is no longer Hermitian. However we may redefine the Hamiltonian in the following way:
\begin{equation}
H^{\rm sys}:=\frac{1}{2}(H_\mathrm{aux}+H^{\dagger}_\mathrm{aux}),
\end{equation}
where $H^{\rm sys}$ is Hermitian now. The effect that $H^{\dagger}_\mathrm{aux}$ has in the system is a redefinition of the orientation on the lattice. So we have a superposition of a lattice orientated in the way of figure~\ref{Operatorsd3} (arrows up and right) and another with arrows down and left. Nevertheless, one can always think in terms of $H_\mathrm{aux}$ for the pictorial image and then use $H^{\rm sys}$ to compute energies and derive equations.

\subsection{Anyon Model}

The theory developed above was done for the general case of qudits. From now on and to be concrete concerning thermal effects, we will focus on the case where $d=3$ (qutrits).
Later on we will be able to extract conclusions for qudits as well. There are still many important aspects  to be  studied about this model and its coupling to a thermal bath. We need to compute the energy gap of the Hamiltonian, i.e. the energy difference between the ground state where the code lies and the excited states which represent the errors. It is also important to calculate the anyon statistics, as long as they are associated with the excitations of a topological system with qutrits.

In $d=3$ the phase factors are $\omega=\r e^{{\rm i}\frac{2\pi}{3}}, \omega^2=\r e^{{\rm i}\frac{4\pi}{3}},\omega^3=1$. We will see for this particular case, how excitations can be created, moved and annihilated.  This will give us the properties of the anyon model which is going to be associated with the group $\mathbb{Z}_3$.

As we did before, we use a notation in which $\sigma_j^z=Z_j$ and $\sigma_j^x=X_j$, except that we use the symbol $\sigma$ to denote errors acting on the system, i.e., bump operators acting because of the coupling to the thermal bath, whereas we shall use  $X,Z$ for the Hamiltonian interactions defined by the vertex and plaquette operators
of $H^{\rm sys}$.

Errors on the system can be expressed in terms of operators $\sigma^x$, $\sigma^z$ or products containing  them,  and acting on each edge $j$ where the qutrits are placed.
And the same goes for $\sigma^z$. To see what is the effect of these errors on the system, we will see how  the ground state changes by applying $\sigma^{x,z}$. We will see that this corresponds to processes in which anyons are created, annihilated or moved throughout the torus.

\noindent Let us see what happens when we bump a qutrit in a position $j$ from the outside and then act with the Hamiltonian $H_\mathrm{aux}$,
$$H_\mathrm{aux}\sigma_j^z\ket{\psi}.$$
Note that every operator of the Hamiltonian commutes with this $\sigma_j^z$ except two $A_s$ operators which share a \emph{leg} with this qubit $j$. But, contrary to the case of $d=2$ there is an orientation defined on the lattice. So, for instance, if an error ($\sigma_j^z$) occurs in a certain vertical edge, one of these $A_s$ (the one below) is defined with an $X_j$, thus:
\begin{equation}
A_s\sigma_j^z\ket{\psi}=\omega^{-1}\sigma_j^zA_s\ket{\psi}=\omega^2\sigma_j^z\ket{\psi},
\end{equation}
but the $A_{s'}$ above the edge is defined with $X^{-1}_j$, then:
\begin{equation}
A_{s'}\sigma_j^z\ket{\psi}=\omega\sigma_j^zA_{s'}\ket{\psi}=\omega\sigma_j^z\ket{\psi}
\end{equation}
Hence, we have two violations of the vertex condition, one with charge $\omega$ and the other with $\omega^2$. This is one of the two types of anyons that we will have in this system, and we shall denote it as an $\omega^2$ --- $\omega$ anyon. It is important to point out that these are only labels to classify the excitations based on the violations of the operator $A_{s}$ (and $B_{p}$). In principle we could classify anyons based on the violation of stabilizers $A_{s}^{-1}$ (and $B_p^{-1}$) that appears in $H_\mathrm{aux}^{\dagger}$. It is just a matter of labeling, the physics is the same.

Now we can act with $\sigma_j^z$ again and obtain the other anyon type called $\omega$ --- $\omega^2$. Actually they could be considered as the same anyon type as before but with opposite orientation. However, it is convenient to define them as two types of anyons as they will have different braiding properties. Moving anyons of the same type around each other will be different from the case of having anyons of different type. Likewise,  it will be necessary to have anyons of different types in order  to have fusion of anyons without annihilation. We shall explain this in the next subsection in more detail.

Note that by acting twice with $\sigma_j^z$ is equivalent to act with $(\sigma_j^z)^{-1}$. Thus, although every error can be expressed in terms of $X$ and $Z$ operators, it will be useful to think sometimes as if we act either with $X,Z$ or $X^{-1},Z^{-1}$. All these arguments are exactly the same in the case of $B_p$ operators and $\sigma^x$ errors. Therefore, we have 4 types of anyons, 2 of plaquette type and 2 of vertex type.

\begin{figure}[h]
\centering
\includegraphics[scale=0.3]{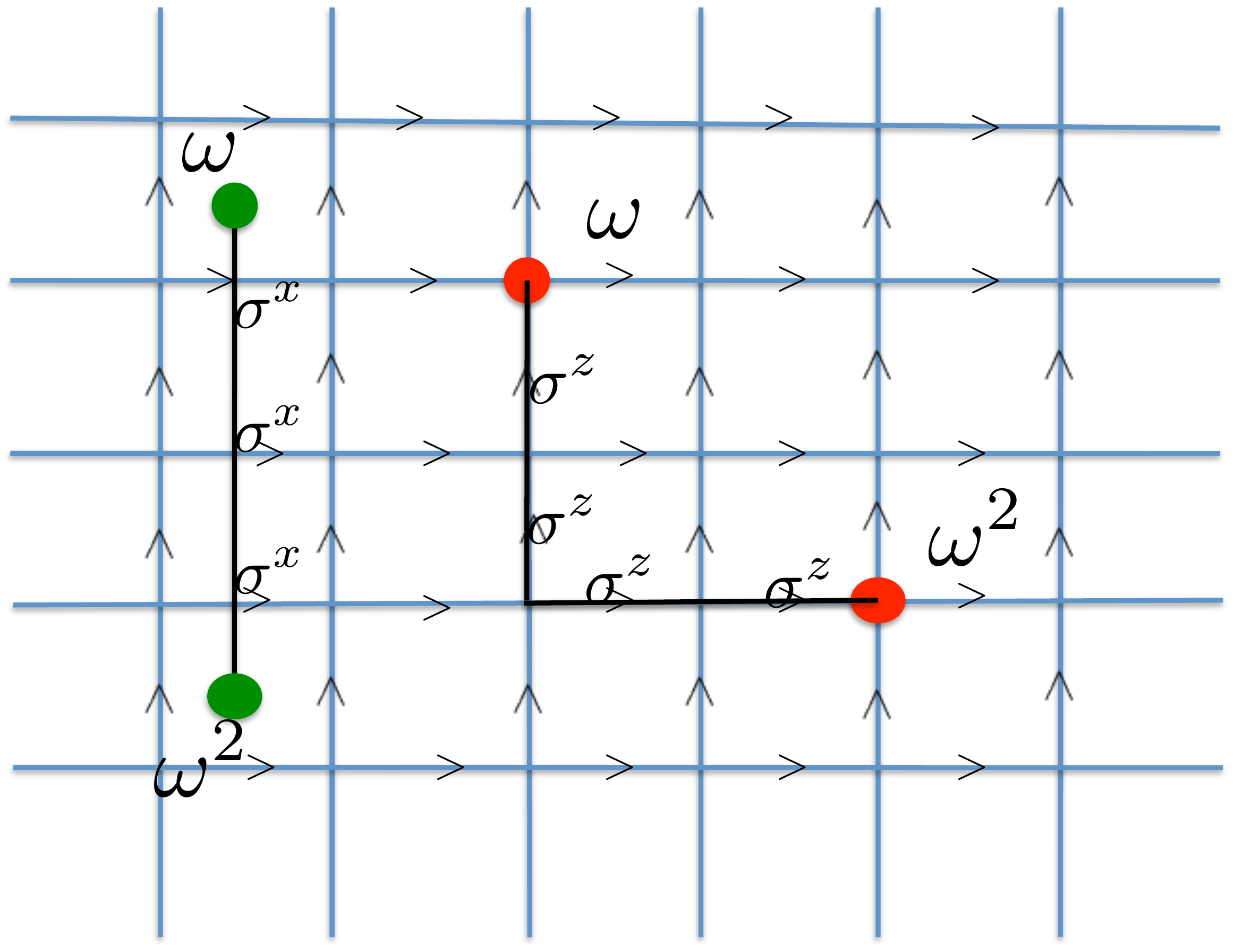}
\caption{Anyons of type $x$ (red) on the direct lattice. Anyons of type $x$ (green) on the dual lattice.}
\label{Anyonsd3}
\end{figure}

\begin{figure}[h]
\centering
\includegraphics[scale=0.3]{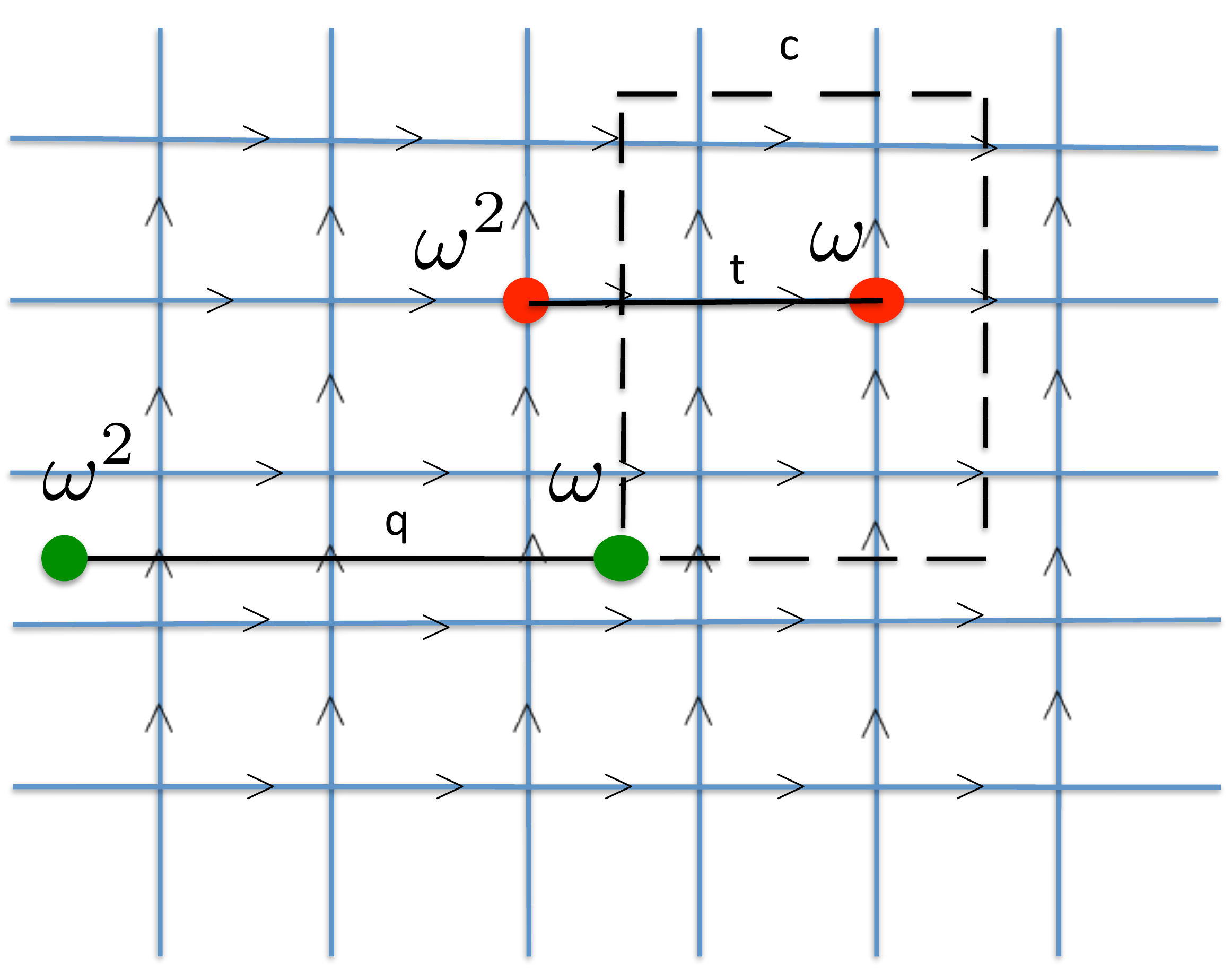}
\caption{Anyons of type Z (red) on the direct lattice attached to a string $t$.
Anyons of type $x$ (green) on the dual lattice associated to a string  $q$. The
$x$-type particle moves around a $z$-type particle on a closed string  $c$.}
\label{Moveanyonsd3}
\end{figure}

Let us study now the braiding of the anyons. We will consider two chains of different type: plaquette anyon and vertex anyon (as in figure  \ref{Anyonsd3}). In this case we get something remarkably different from the $d=2$ case. Now it is not the same to let one anyon still and move the other around it than do it the other way around.
Thus,  let us  move particles around each other. For example, let us move
an $x$-type particle around a $z$-type particle (see figure ~\ref{Moveanyonsd3}).
Then,
\[
  |\Psi_{\mbox{\scriptsize initial}}\rangle\, =\, S^z(t)\,|\psi^x(q)\rangle
  \ ,\qquad\quad
  |\Psi_{\mbox{\scriptsize final}}\rangle\, =\,
  S^x(c)\,S^z(t)\,|\psi^x(q)\rangle\ =\
  \omega^2|\Psi_{\mbox{\scriptsize initial}}\rangle,
\]
because $S^x(c)$ and $S^z(t)$ cross each other just on one qutrit satisfying the relation $$XZ=\omega^2 ZX$$ and $S^x(c)|\psi^x(q)\rangle=|\psi^x(q)\rangle$.
We see that the global wave function, i.e. the state of the entire system,
acquires the phase factor $\omega^2$.
Nonetheless, if the operation is the opposite, that is, if we move a $z$-type particle around a $x$-type particle then:
\[
  |\Psi_{\mbox{\scriptsize initial}}\rangle\, =\, S^x(q)\,|\psi^z(t)\rangle
  \ ,\qquad\quad
  |\Psi_{\mbox{\scriptsize final}}\rangle\, =\,
  S^z(c)\,S^x(q)\,|\psi^z(t)\rangle\ =\
  \omega|\Psi_{\mbox{\scriptsize initial}}\rangle,
\]
since $S^x(q)$ and $S^z(c)$ cross each other just on one qutrit again satisfying the relation:
\begin{equation} ZX=\omega XZ\end{equation}
and $S^z(c)|\psi^z(t)\rangle=|\psi^z(t)\rangle$.
We see that the global wave function
acquires now the phase factor $\omega$.

Therefore, we arrive at a very important novelty for qutrits that is different from when we dealt with qubits regarding two aspects:
\begin{enumerate}
\item The phase that the anyon picks up is different from $-1$.
\item The phase depends on the orientation in which the braiding close path is traversed.
\end{enumerate}

\subsection{New Anyon Energy Processes}

First of all, let us look at the gap of the Hamiltonian.
We will reach our first excited state by applying a $\sigma^z$ or $\sigma^x$ operator to the ground state. Let us see which is the energy difference between the ground state and the first excited state. Remember that $2H^{\rm sys}=H_\mathrm{aux}+H_\mathrm{aux}^{\dagger}$. We denote $P$ and $S$ the number of plaquette and vertex operators respectively, with $P+S=N$ the number of qutrits in the lattice,  and $\{l,l'\}$ are the adjacent vertices of the site of a qutrit $j$:
\begin{eqnarray}
H\ket{\psi}&=&\frac{1}{2}\{-\sum_s A_s \,-\, \sum_p B_p+h.c\}\ket{\psi}=-(P+S)\ket{\psi}\\
H\sigma^z_{j}\ket{\psi}&=&\frac{1}{2}\{-\sum_s A_s \,-\, \sum_p B_p+h.c\}\sigma^z_{j}\ket{\psi}=-(P+S-2)\sigma^z_j\ket{\psi}-\frac{1}{2}(A_l\sigma_j^z\ket{\psi}+\nonumber\\
&-&A_{l'}\sigma_j^z\ket{\psi}+A_l^{\dagger}\sigma_j^z\ket{\psi}-A_{l'}^{\dagger}\sigma_j^z\ket{\psi})=-(P+S-2)\sigma^z_j\ket{\psi}-\omega^2\sigma_j^z\ket{\psi}-\omega\sigma_j^zA_{l'}^{\dagger}\ket{\psi}=\nonumber\\
&=&-(P+S-2+\omega+\omega^2)\sigma^z_{j}\ket{\psi}=-(P+S-2+2\cos{\frac{2\pi}{3}})\sigma^z_j\ket{\psi}=\nonumber\\
&=&-(P+S-3)\sigma^z_j\ket{\psi}.\nonumber
\end{eqnarray}
Thus, the energy difference is
\begin{equation*}
\Delta E=3.
\end{equation*}
The action of $\sigma^x$ produces the same energy increment but we have to do the commutation with the operators $B_p$.

This calculation can be easily extended to the case of \emph{qudits} with arbitrary $d$, obtaining the gap equation
\begin{equation}
\Delta E=\Delta_d=2\left(1-\cos{\frac{2\pi}{d}}\right).
\end{equation}

\begin{figure}[h]
\centering
\includegraphics[scale=0.3]{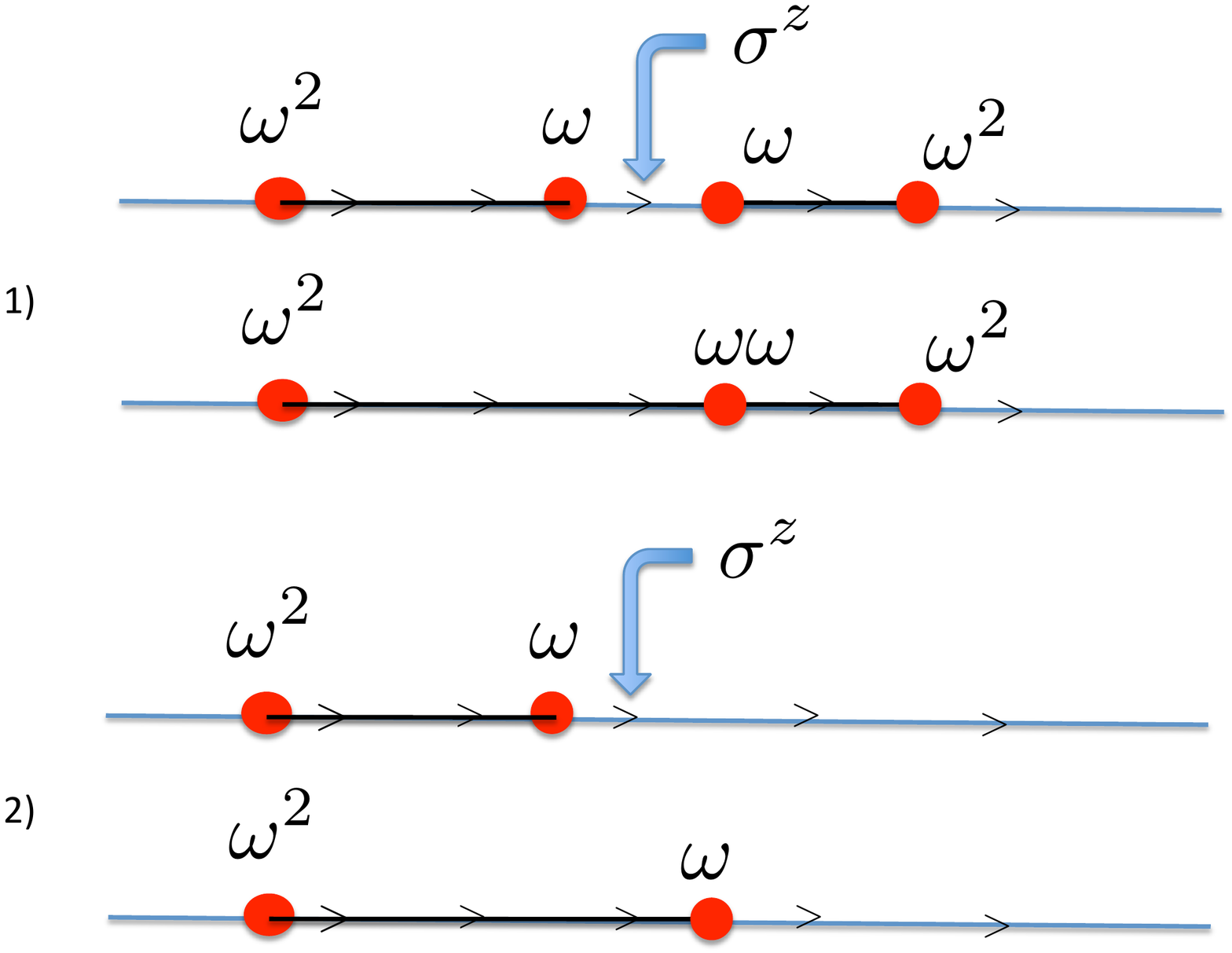}
\caption{1) Fusion of anyons (ending tied, not annihilated). 2) Movement of an anyon. We plot just one dimension as long as the rest of the lattice is irrelevant, i.e, the process is the same everywhere.}
\label{Anyproc}
\end{figure}

Note that there is a reduction of the energy gap for $d=3$ in comparison with the case of qubits, where it was 4. It is also important to point out that if we act again on the same bond of the lattice with $(\sigma^z)^{-1}$, there would be an energy reduction of the same amount of energy. Moreover, if at the endpoint of an anyon $\omega$ --- $\omega^2$ we act with $\sigma^z$ we obtain the same pair of anyons again,  and same energy,  but longer (see figure~\ref{Anyproc}.2). In this process the energy is preserved $\Delta E=0$. This means that there is no energy exchange between the thermal bath and the system. We can understand the process as a diffusion of the anyon with no energy cost. In analogy to the case $d=2$, this is what is called moving an anyon. It is also important to remark that still for qutrits, all process that involve moving a simple pair of anyons have no energy cost.

Until here, there is a complete analogy with the case of $d=2$. But we are going to see now a process that only occurs in $d>2$. Imagine that there have been two excitations on the system, and two anyons of opposite orientation have been created. Moreover, they are separated by just one vertex operator. The situation is plotted in figure~\ref{Anyproc}.1.

\noindent Imagine that we act now with a $\sigma^z$ on the bond, which is error free,  that links the anyons $\omega^2$ --- $\omega$ and $\omega$ --- $\omega^2$ (opposite orientation). Let us analyze the energy process.
\begin{eqnarray}
H\ket{\psi'}&=&-\frac{1}{2}(-\sum_s A_s \,-\, \sum_p B_p)\ket{\psi'}+h.c.=-(P+S-6)\ket{\psi},\\
H\sigma^z_j\ket{\psi'}&=&-\frac{1}{2}(-\sum_s A_s \,-\, \sum_p B_p)\sigma^z_j\ket{\psi'}+h.c=-\frac{1}{2}(P+S-\omega-\omega-\omega^2\omega^2)\sigma^z\ket{\psi'}-\nonumber\\
&-&\frac{1}{2}(P+S-\omega^2-\omega^2-\omega\omega)\sigma^z\ket{\psi'}=-(P+S-6+\frac{3}{2})\sigma^z\ket{\psi'},\nonumber
\end{eqnarray}
so, the energy difference is
\begin{equation*}
\Delta E=-3/2.
\end{equation*}
What has occurred is that two anyons have been tied together, but not annihilated. This process lowers the energy of the system in a smaller amount than the process of annihilation. If in this situation we would act with a $(\sigma^z_j)^{-1}$ on the point where the two pair of anyons are tied together, the two anyons would split apart, and this process would cost energy $\Delta E=3/2$. This could be analyzed exactly the same way with $\sigma^x$ errors and $B_p$ operators.\\

It is remarkable that this phenomenon cannot happen in $d=2$, as in $d=2$ the product $\omega\omega=(-1)(-1)=1$. Therefore, $d=3$ is the first non trivial case to have processes like these in a toric code with qudits.

\subsection{Master Equation for Topological Qutrits}

As we have seen, all these processes are generated by the action of operators $\sigma^z$, $(\sigma^z)^2$ and $\sigma^x$, $(\sigma^x)^2$; as in this case, the square of the Pauli operators are their Hermitian conjugate. Nevertheless, the energy exchange depends on the situation of the system when we bump it with the thermal bath from outside. Before writing the master equation that describes the dynamics of the system, it will be useful to distinguish between these situations by local projectors. The answer to the question whether this is possible or not in this case is not trivial. However, we show that it is possible to classify into groups of processes that have the same energy gain from the bath. Furthermore, they could be distinguished by certain projection operators that only involve two adjacent vertex or plaquette operators.

\noindent We arrive at the following classification:

\begin{eqnarray}
1 & \text{------} & 1 \hspace{1.7cm} P^j_{++}=4\mathcal{A}^{(1)}_{\alpha=+1}(s)\mathcal{A}^{(1)}_{\alpha=-1}(s)\mathcal{A}^{(1)}_{\alpha=+1}(s')\mathcal{A}^{(1)}_{\alpha=-1}(s)\nonumber\\
\hline
\omega & \text{------} & 1 \hspace{1.7cm} P^j_{+(1)}=8\mathcal{A}^{(2)}_{\alpha=0}(s,s')\mathcal{A}^{(2)}_{\alpha=+1}(s,s')\Delta \mathcal{A}(s,s')\Delta \mathcal{A}^{\dagger}(s,s')\mathcal{A}^{(1)}_{\alpha=-1}(s)\mathcal{A}^{\dagger(1)}_{\alpha=-1}(s)\nonumber\\
1 & \text{------} & \omega \hspace{1.7cm} P^j_{+(2)}=8\mathcal{A}^{(2)}_{\alpha=0}(s,s')\mathcal{A}^{(2)}_{\alpha=+1}(s,s')\Delta \mathcal{A}(s,s')\Delta \mathcal{A}^{\dagger}(s,s')\mathcal{A}^{(1)}_{\alpha=-1}(s)\mathcal{A}^{\dagger(1)}_{\alpha=-1}(s)\nonumber\\
\hline
\omega^2 & \text{------} & 1 \hspace{1.7cm}P^j_{0(1)}=8\mathcal{A}^{(2)}_{\alpha=0}(s,s')\mathcal{A}^{(2)}_{\alpha=-1}(s,s')\Delta \mathcal{A}(s,s')\Delta \mathcal{A}^{\dagger}(s,s')\mathcal{A}^{(1)}_{\alpha=+1}(s')\mathcal{A}^{\dagger(1)}_{\alpha=+1}(s')\nonumber\\
1 & \text{------} & \omega^2 \hspace{1.7cm}P^j_{0(2)}=8\mathcal{A}^{(2)}_{\alpha=0}(s,s')\mathcal{A}^{(2)}_{\alpha=-1}(s,s')\Delta \mathcal{A}(s,s')\Delta \mathcal{A}^{\dagger}(s,s')\mathcal{A}^{(1)}_{\alpha=+1}(s)\mathcal{A}^{\dagger(1)}_{\alpha=+1}(s)\nonumber\\
\omega & \text{------} & \omega \hspace{1.7cm}P^j_{0(3)}=8\mathcal{A}^{(2)}_{\alpha=0}(s,s')\mathcal{A}^{(2)}_{\alpha=-1}(s,s')\mathcal{A}^{(1)}_{\alpha=+1}(s)\mathcal{A}^{\dagger(1)}_{\alpha=+1}(s)\mathcal{A}^{(1)}_{\alpha=+1}(s')\mathcal{A}^{\dagger(1)}_{\alpha=+1}(s')\nonumber\\
\hline
\omega^2 & \text{------} & \omega \hspace{1.7cm}P^j_{-(1)}=8\mathcal{A}^{(2)}_{\alpha=+1}(s,s')\mathcal{A}^{(2)}_{\alpha=-1}(s,s')\Delta \mathcal{A}(s,s')\Delta \mathcal{A}^{\dagger}(s,s')\mathcal{A}^{(1)}_{\alpha=+1}(s')\mathcal{A}^{\dagger(1)}_{\alpha=+1}(s')\nonumber\\
\omega & \text{------} & \omega^2 \hspace{1.7cm}P^j_{-(2)}=8\mathcal{A}^{(2)}_{\alpha=+1}(s,s')\mathcal{A}^{(2)}_{\alpha=-1}(s,s')\Delta \mathcal{A}(s,s')\Delta \mathcal{A}^{\dagger}(s,s')\mathcal{A}^{(1)}_{\alpha=+1}(s)\mathcal{A}^{\dagger(1)}_{\alpha=+1}(s)\nonumber\\
\hline
\omega^2 & \text{------} & \omega^2 \hspace{1.7cm}P^j_{--}=8\mathcal{A}^{(2)}_{\alpha=0}(s,s')\mathcal{A}^{(2)}_{\alpha=+1}(s,s')\mathcal{A}^{(1)}_{\alpha=-1}(s)\mathcal{A}^{\dagger(1)}_{\alpha=-1}(s)\mathcal{A}^{(1)}_{\alpha=-1}(s')\mathcal{A}^{\dagger(1)}_{\alpha=-1}(s')\nonumber\\
\label{opP}
\end{eqnarray}

\begin{figure}[h]
\centering
\includegraphics[scale=0.4]{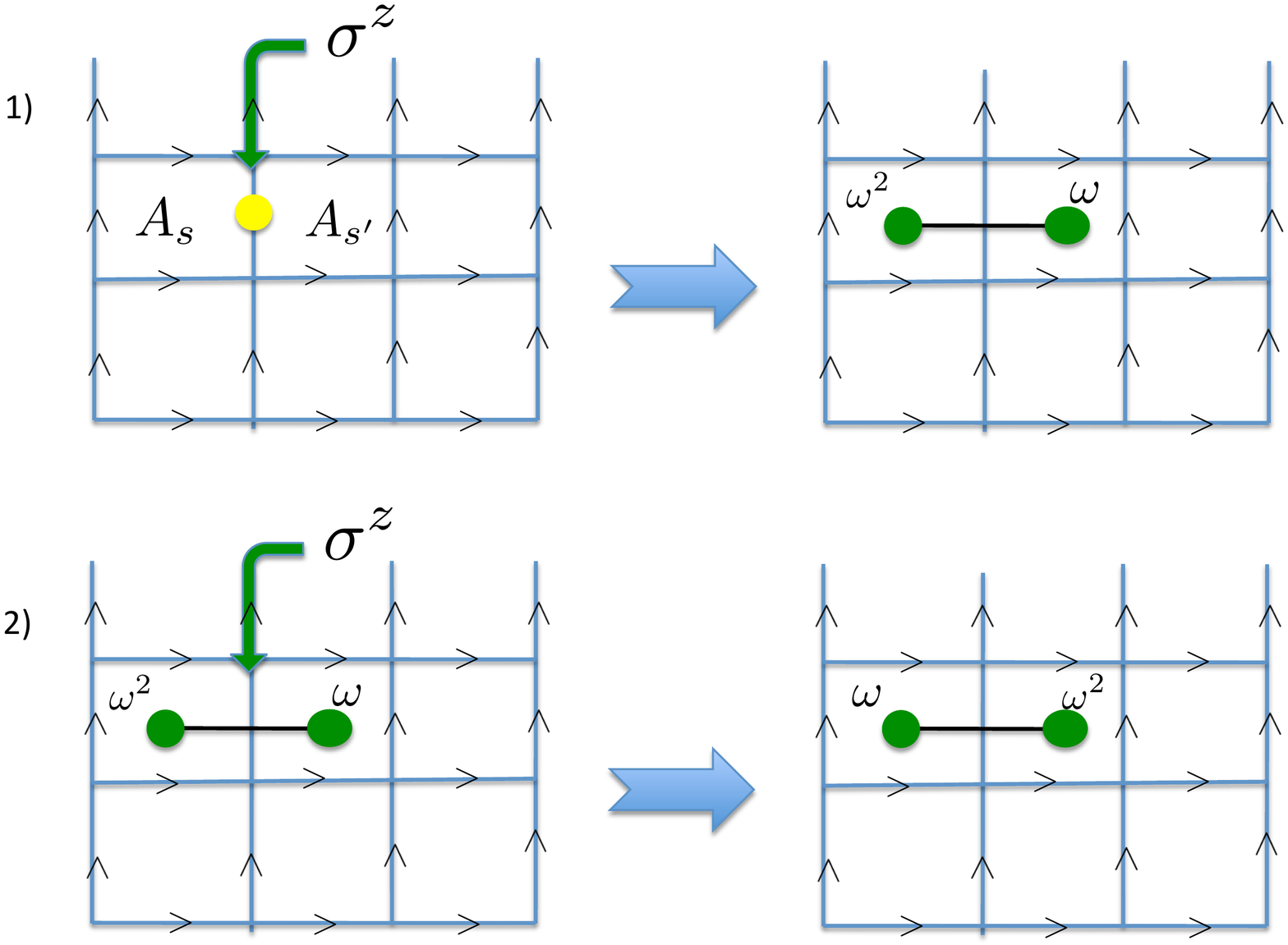}
\caption{1) Initial state $1~$---$~1$ $\Longrightarrow$ Final state $\omega^2~$---$~\omega$~~~~~2) Initial state $\omega^2~$---$~\omega$ $\Longrightarrow$ Final state $\omega~$---$~\omega^2$.~~~~This is an example of what happens to the topological charges when there is a bump from the thermal bath outside. The first one the energy gain is $\Delta E=3$. The second one $\Delta E=0$.}
\label{Charges}
\end{figure}

In this table we have represented all combinations of two adjacent topological charges.
In the first column: we depict a representation of the different types of anyons, with two topological charges attached at their ends and linked by a dash.
Correspondingly, all these anyons have an intrinsic orientation.
At the left side of the dash there is the eigenvalue of the operator $A_s$ and at the right side, the eigenvalue of the adjacent operator $A_s'$. A physical qutrit $j$ would be in the middle of the dash (see an example at figure~\ref{Charges}).
In the second column: we write the projector that gives 1 for that situation and 0 for the others.

Here we have defined the following operators in order to simplify the notation:
\begin{eqnarray}
{\mathcal{A}}^{(1)}_{\alpha=0,+1,-1}(s)&:=&(1-\omega^{\alpha}A_s),\nonumber\\
{\mathcal{A}}^{(2)}_{\alpha=0,+1,-1}(s,s')&:=&(1-\omega^{\alpha}A_sA_s')\nonumber,\\
\Delta \mathcal{A}(s,s')&:=&{\mathcal{A}}^{(1)}_{\alpha=0}(s')-{\mathcal{A}}^{(1)}_{\alpha=0}(s),\nonumber
\end{eqnarray}
where $s$ and $s'$ are the two vertex surrounding the qutrit $j$. The index $\alpha$ takes values on the exponent of the phases $\omega $ that appear from the braiding processes. These projectors tell us which are the charges of the system that surround a certain qutrit. That is  why they are local projectors. Moreover it is easy to verify that they form a set of orthogonal projectors:
\begin{eqnarray}
\sum_{\alpha}P^j_{\alpha}&=&\mathds{1},\nonumber\\
P^j_{\alpha}&=&P_{\alpha}^{j\dagger},\nonumber\\
(P^j_{\alpha})^2&=&P^j_{\alpha}.\nonumber
\end{eqnarray}
\indent As we have already explained, we classify the situation of the system in terms of the charges according to the eingenvalues of the operators $A_s$ associated with the part of the Hamiltonian $H_\mathrm{aux}$. One could do the same thing for $A^{-1}_s$, but the situation of the system will be the same independently of the label we assign to them. So these projectors can discriminate perfectly between eigenstates of the Hamiltonian $H^{\rm sys}$.

Now, given a certain state of the system $\ket{\psi'}$, by applying these projectors we can figure out which situation we have. This means that if an operator $\sigma^z$ or $\sigma^x$ (or their Hermitian conjugate) is going to act on our system, we will know which energy process is bound to happen. Based on this, and studying the different situations that we can encounter, one can define a set of operators that tells us whether an anyon has been moved, created, annihilated or fused when we apply the generalized Pauli operators (as we did in figure~\ref{Anyproc}). This is done by analyzing the initial and the final state after the action of a bump operator and seeing which would be the energy after and before the process, as shown in figure ~\ref{Charges}. Therefore,  we have:

\begin{eqnarray}
a^{(1)\dagger}_j&:=&\sigma^z_jP^j_{++}+(\sigma^{z}_j)^{-1}P^j_{++}, \nonumber\\
a^{(1)}_j&:=&(\sigma^{z}_j)^{-1}P^j_{-(1)}+\sigma^z_jP^j_{-(2)},\nonumber\\
a^{(2)\dagger}_j&:=&(\sigma^{z}_j)^{-1}P^j_{+(1)}+\sigma^z_jP^j_{+(2)}+\sigma^z_jP^j_{0(1)}(\sigma^{z}_j)^{-1}P^j_{0(2)},\\
a^{(2)}_j&:=&(\sigma^{z}_j)^{-1}P^j_{0(3)}+\sigma_j^zP^j_{0(3)}+\sigma^z_jP^j_{--}+(\sigma^{z}_j)^{-1}P^j_{--},\nonumber\\
a^0_j&:=&\sigma^z_jP^j_{+(1)}+(\sigma^{z}_j)^{-1}P^j_{+(2)}+\sigma^z_jP^j_{0(2)}(\sigma^{z}_j)^{-1}P^j_{0(1)}+\sigma^z_jP^j_{-(1)}+(\sigma^{z}_j)^{-1}P^j_{-(2)}.\nonumber
\label{opa}
\end{eqnarray}
Here the upper-indices of operators $a_j$ are related to the energy cost of the process:
\begin{itemize}
\item $a^{(1)\dagger}_j$ creates a pair of anyons of $z$-type and $a^{(1)}_j$ annihilates it. The energy cost is $\Delta E=3$.
\item $a^{(2)\dagger}_j$ and $a^{(2)}_j$ are related to the process of fusion or  separation, respectively, of anyons as in figure \ref{Anyproc}.1 and also to the process of creation (and annihilation) of a pair of anyons tied to a previous pair. The energy cost is $\Delta E=\frac{3}{2}$.
\item $a^0_j$ moves anyons and also it can invert the orientation of a pair of anyons (as in figure~\ref{Charges}.2). There is no energy cost in these processes.
\end{itemize}

For the plaquette operators $B_p$ we proceed in the same way obtaining a similar result. The corresponding local projectors that we denote as $R_j$ are built analogously just by changing $A_s$ for $B_p$, where $p$ and $p'$ are the adjacent plaquettes to the qutrit $j$. Then the operators which describe the analogous process for $x$-type anyons are:

\begin{eqnarray}
b^{(1)\dagger}_j&:=&\sigma^x_jR^j_{++}+(\sigma^{x}_j)^{-1}R^j_{++}, \nonumber\\
b^{(1)}_j&:=&(\sigma^{x}_j)^{-1}R^j_{-(1)}+\sigma^x_jR^j_{-(2)},\nonumber\\
b^{(2)\dagger}_j&:=&(\sigma^{x}_j)^{-1}R^j_{+(1)}+\sigma^x_jR^j_{+(2)}+\sigma^x_jR^j_{0(1)}(\sigma^{x}_j)^{-1}R^j_{0(2)},\\
b^{(2)}_j&:=&(\sigma^{x}_j)^{-1}R^j_{0(3)}+\sigma^x_jR^j_{0(3)}+\sigma^x_jR^j_{--}+(\sigma^{x}_j)^{-1}R^j_{--},\nonumber\\
b^0_j&:=&\sigma^x_jR^j_{+(1)}+(\sigma^{x}_j)^{-1}R^j_{+(2)}+\sigma^x_jR^j_{0(2)}(\sigma^{x}_j)^{-1}R^j_{0(1)}+\sigma^x_jR^j_{-(1)}+(\sigma^{x}_j)^{-1}R^j_{-(2)}\nonumber.
\label{opb}
\end{eqnarray}
Some of these operators are associated to more than one projector, unlike for qubits. That is because for 3-level systems, the possibilities for different excitations scenarios have grown significantly. 

As we have seen in the previous section, these operators arise naturally as the Fourier transform of the interaction Hamiltonian when a thermal bath is weakly coupled with our system,

\begin{equation}
\r e^{{\rm i}tH^{\mathrm{sys}}}\, S_\alpha\, \r e^{-{\rm i}tH^{\mathrm{sys}}} = \sum_\omega S_\alpha(\omega)\, \r e^{-{\rm i}\omega t}.
\label{salpha}
\end{equation}
In this case the interaction  Hamiltonian will be of the form:
\begin{equation}
V=\sum_{\alpha}S_{\alpha}\otimes f_{\alpha}=\sum_j\sigma^z_j\otimes f^z_j+(\sigma^z_j)^{-1}\otimes (f^z_j)^{\dagger}+\sigma^x_j\otimes f^x_j+(\sigma^x_j)^{-1}\otimes (f^x_j)^{\dagger},
\end{equation}
and it is quite important to remark that there are only 3 Bohr frequencies this time, $\omega=0,\pm\frac{3}{2},\pm3$.

We can check that the dynamical operators obtained are indeed compatible with this interaction potential as $\sum_{\alpha}S_{\alpha}=\sum_{\alpha}S_{\alpha}(\omega)$. In our case, it is trivial to check:
 \begin{eqnarray}
 \sigma^z_j+(\sigma^z_j)^{-1}=\sum_n a^n_j,\nonumber\\
 \sigma^x_j+(\sigma^x_j)^{-1}=\sum_n b^n_j\nonumber\\
 \nonumber
 \end{eqnarray}
 with $n=0,1,2$, using equations \eqref{opa} and \eqref{opb}.

Moreover, $[H,a^n]\propto a^n$, based on the fact that $H^{\rm sys}$ is made of stabilizers which at most introduces a phase when they are applied to states $a^i\ket{\phi}$. Thus, $A_s(B_p)a^i\ket{\phi}\propto a^i\ket{\phi}$ and $a^iA_s(B_p)\ket{\phi}\propto a^i\ket{\phi}$ therefore $[H,a^i]\propto a^i, ~\forall a^i$ a dynamical operator of our system.
With this proviso, the Davies generator turns out to be given by:
\begin{equation}
\frac{dX}{dt} = \c G(X) = \r i \delta(X) + \c L(X),
\label{reddyn}
\end{equation}
with
\begin{eqnarray}
\delta(X)&=& [H^{\mathrm{sys}},X]=\tfrac{1}{2}[H_{\mathrm{aux}}+H_{\mathrm{aux}}^\dagger,X]\nonumber,\\
\mathcal{L}(X)&=&\mathcal{L}^z(X)+\mathcal{L}^x(X),\nonumber\\
\mathcal{L}^x(X)&=&\sum_j\frac{1}{2}R(3)\{(-b^{(1)\dagger}_jb^{(1)}_jX-Xb^{(1)\dagger}_jb^{(1)}_j+2b^{(1)\dagger}_jXb^{(1)}_j)+\r e^{-3\beta}(-b^{(1)}_jb^{(1)\dagger}_jX-Xb^{(1)}_jb^{(1)\dagger}_j+\nonumber\\
&+&2b^{(1)}_jXb^{(1)\dagger}_j\})+\frac{1}{2}R(3/2)\{(-b^{(2)\dagger}_jb^{(2)}_jX-Xb^{(2)\dagger}_jb^{(2)}_j+2b^{(2)\dagger}_jXb^{(2)}_j)+\r e^{-\frac{3}{2}\beta}(-b^{(2)}_jb^{(2)\dagger}_jX-\nonumber\\
&-&Xb^{(2)}_jb^{(2)\dagger}_j+2b^{(2)}_jXb^{(2)\dagger}_j)\}-\frac{1}{2}R(0)[b^0_j,[b^0_j,X]],\nonumber\\
\mathcal{L}^z(X)&=&\sum_j\frac{1}{2}R(3)\{(-a^{(1)\dagger}_ja^{(1)}_jX-Xa^{(1)\dagger}_ja^{(1)}_j+2a^{(1)\dagger}_jXa^{(1)}_j)+\r e^{-3\beta}(-a^{(1)}_ja^{(1)\dagger}_jX-Xa^{(1)}_ja^{(1)\dagger}_j+\nonumber\\
&+&2a^{(1)}_jXa^{(1)\dagger}_j\})+\frac{1}{2}R(3/2)\{(-a^{(2)\dagger}_ja^{(2)}_jX-Xa^{(2)\dagger}_ja^{(2)}_j+2a^{(2)\dagger}_jXa^{(2)}_j)+\r e^{-\frac{3}{2}\beta}(-a^{(2)}_ja^{(2)\dagger}_jX-\nonumber\\
&-&Xa^{(2)}_ja^{(2)\dagger}_j+2a^{(2)}_jXa^{(2)\dagger}_j)\}-\frac{1}{2}R(0)[a^0_j,[a^0_j,X]].\label{Operadores3ab}
\end{eqnarray}

\subsection{Topological Order}

Similarly to the case of qubits, we will study the evolution of the expectation value $\bra{{\rm GS}}X_c\ket{{\rm GS}}$, where $X_c$ is the tensor product of $\sigma^x$ generalized Pauli operators $(d=3)$ along a non-contractible loop, and $\ket{{\rm GS}}$ denotes a certain ground state in the stabilizer subspace; namely a superposition of the degenerate states in the ground state manifold of $H^{{\rm sys}}$.

In the weak coupling limit, the master equation that describes the dynamics of this quantity is:
\begin{equation}
\frac{dX_c(t)}{dt}=\r i[H^{\mathrm{sys}},X_c(t)]+\mathcal{L}[X_c(t)]. \label{masterX}
\end{equation}
In order to simplify the calculation we remove the free evolution part of the equation
\begin{equation}
\tilde{X}_c(t)=\r e^{-\r i H^{\mathrm{sys}}t}X_c(t)\r e^{\r i H^{\mathrm{sys}}t}~\Longrightarrow
\frac{d\tilde{X}_c(t)}{dt}=\mathcal{L}[\tilde{X}_c(t)],\label{masterXsim}
\end{equation}
being both the dissipator $\mathcal{L}$ and the mean value $\bra{{\rm GS}}X_c\ket{{\rm GS}}$ invariant under this transformation.

\subsection{Short time regime}

In the short time regime, we can approximate $\tilde{X}_c(t)\simeq(1+t\mathcal{L})X_c$; here we denote $X_c:=X_c(0)$. Thus, the evolution of $\bra{{\rm GS}}X_c(t)\ket{{\rm GS}}$ is
\begin{equation}
\langle\tilde{X}_c(t)\rangle\simeq\bra{{\rm GS}}X_c\ket{{\rm GS}}+t\bra{{\rm GS}}\mathcal{L}(X_c)\ket{{\rm GS}}.
\end{equation}

We need to calculate $\bra{{\rm GS}}\mathcal{L}(X_c)\ket{{\rm GS}}$, with $\mathcal{L}(X_c)=\mathcal{L}^x(X_c)+\mathcal{L}^z(X_c)$. This calculation is made in Appendix \ref{App-TOqutrits}, obtaining
\begin{equation}
\bra{{\rm GS}}\mathcal{L}(X_c)\ket{{\rm GS}}=-\frac{\Delta}{2}R(\Delta)\r e^{-\Delta\beta}|c|\bra{{\rm GS}}X_c\ket{{\rm GS}}.\\
\end{equation}
Hence, we can define $\Gamma:=\frac{\Delta}{2}R(\Delta)\r e^{-\Delta\beta}|c|$ as the initial decay rate of the system. For qutrits $\Delta=3$ while for qubits (see Eq. \eqref{Gammaqubit}) we have obtained an analogous expression but with $\Delta=4$ instead.

This result can be generalized for the case of \emph{qudits} with arbitrary $d$. We have already seen, that at short times, only the creation of anyons contributes to the decay of topological order. The free diffusion of anyons and the fusion processes among them will not appear as they are second order processes in time. However, as we increase $d$ there are more types of anyons with different energies. Moreover, a pair of anyons should always be compatible with the conditions $\prod_s A_s=1$\, and $\prod_p B_p=1$. That means that the possible types of anyons with different energies are of the form $\omega^n-\omega^{d-n}$ with $n=1,...,\lfloor{\frac{d}{2}}\rfloor$, and respective energies $\Delta_n=2(1-\cos{\frac{2\pi n}{d}})$. Note that $n=1$ refers to the lowest energy pair of anyons, i.e. the energy gap of the Hamiltonian. Thus, the initial decay rate has to be the sum of all these contributions:
\begin{equation}
\Gamma_d=\sum_{n=1}^{\lfloor{\frac{d}{2}}\rfloor}\frac{\Delta_n}{2}|c|R(\Delta_n)\text{e}^{-\Delta_n\beta}.
\label{Rated}
\end{equation}
It is important to point out that in the case of qudits, an analogous expression for the interaction with the environment to \eqref{V} involves $S_{\alpha}=\sigma^x,(\sigma^x)^2,...,(\sigma^x)^{d-1},\sigma^z,(\sigma^z)^2,...,(\sigma^z)^{d-1}$. All non trivial powers of $\sigma^x$ and $\sigma^z$ are included to allow for excitations of physical \emph{qudits} from one level to another, at first order in time.

Using Eq. \eqref{Rated} it will be possible to establish a crossover temperature $T_c$ as the limit for which the initial decay rate $\Gamma$  will be larger for qubits than for qudits. For the sake of comparison we take $R(\Delta_n)$ the same for qubits and {qudits}. This is reasonable since $\Delta_n$ are of the same order, and $R(\Delta_n)$ are the Fourier transforms of the bath coupling that induces the excitations on the physical qudits. Thus, we set up the condition $\Gamma_d(T_c):=\Gamma_2(T_c)$.  Using Eq. \eqref{Rated} we arrive at the following expression:
\begin{eqnarray}
4=\sum_{n=1}^{\lfloor{\frac{d}{2}}\rfloor}\Delta_n\text{e}^{-(\Delta_n-4)\beta}>\sum_{n=1}^{\lfloor{\frac{d}{2}}\rfloor}\Delta_n,
\end{eqnarray}
as $\Delta_n<4$ for $d>2$,  $\forall n$.
Therefore, this equation only has a solution for such values of $d$ satisfying $\sum_{n=1}^{\lfloor{\frac{d}{2}}\rfloor}\Delta_n<4$. But this is only true for $d=3$. Thus, there exits only such a $T_c$ for \emph{qutrits}. For other values of $d$, the initial decay rate for qudits will always be larger than for qubits. This happens as $\sum_n\Delta_n$ increases almost linearly with $d$, and $d=3$ is the only case when this quantity is smaller than 4, i.e., the gap in the case of qubits. Let us  now compute $T_c$ for qutrits:

\begin{equation}
3E_0\r e^{-3E_0\beta_c}=4E_0\r e^{-4E_0\beta_c},
\end{equation}
with $E_0$ the natural energy unit of the system. This leads to the following crossover temperature,
\begin{equation}
T_c=\frac{E_0}{k_{\rm B}\ln\frac{4}{3}}.
\end{equation}
The meaning of this temperature is the following. Above this temperature $T_c$, the initial decay rate for qutrits is smaller than for qubits, something that makes qutrits better in this comparison. For $E_0\sim100$kHz used in the proposal of a Rydberg quantum simulator \cite{Markus}  for the operators of the 2-D Toric Code, we obtain an estimate of $T_c\sim20\mu K$.

\noindent In addition, it could be computed a $T_c$ comparing systems with $d$ odd and $(d-1)$ even. There is always a temperature above which the system of \emph{qudits} with $d$ odd has a smaller initial decay rate than the previous $(d-1)$ even.

It is also important to point out that $\Gamma$ is only the initial decay rate. It is possible that the dynamics of anyons, with free diffusion etc.,  play an important role in the loss of topological order. Beyond short times, our conjecture is that the new processes that appears in the case of qutrits, i.e. fusion of anyons which end tied up, will be an obstacle for the free diffusion of anyons. This would represent an improvement for the stability of the generalized toric code in some intermediate time regime for this is the cause of the loss of topological order in the system.

\subsection{Long-Time Regime}

Now we want to study the master equation \eqref{masterX} in the opposite time regime. We are interested in the fate of the non-local order parameter we are using to describe
the topological order in a system of qudits in a generalized toric code. We conjecture that the final state will be given by a thermal Gibbs state.
To show that our observable for the order parameter  $\langle X_c\rangle$ approaches to the expectation value of $X_c$ in the Gibbs state for times long enough,
we resort again to the condition (\ref{ergodicity}). In the generalized case, it reads as follows
\begin{equation}\label{sponhqdit}
\{\sigma_x,\sigma_x^2,\ldots,\sigma_x^{d-1},\sigma_z,\sigma_x^2,\ldots,\sigma_z^{d-1}\}'=\mathds{C1},\quad \text{ for any } d.
\end{equation}
This is due to the fact that if some generic operator, say $A$, commutes with every element of the set $\{\sigma_x,\sigma_x^2,\ldots,\sigma_x^{d-1},\sigma_z,\sigma_x^2,\ldots,\sigma_z^{d-1}\}$, so does with every element of the $d$--Pauli group. This follows from the \mbox{Jacobi} identity and the fact that $\sigma_z\sigma_x=\omega\sigma_x\sigma_z$. Therefore given the irreducibility of the computational representation the $d-$Pauli group (the technical details of this proof are given in Appendix \ref{App-irreducibility}) the condition \eqref{sponhqdit} holds. 

With this result, we may obtain the behaviour in the long time regime

\begin{equation}
\langle X_c(t\rightarrow\infty)\rangle=\mathrm{Tr}(X_c\rho(t\rightarrow\infty))=\frac{1}{Z}\sum_i\text{e}^{-\beta\lambda_i}\bra{\psi_i}X_c\ket{\psi_i}=0,
\end{equation}
which implies that the topological order is also destroyed for qudits in the generalized toric code when times of interaction with a thermal bath are long enough.

Now, let us summarize and combine the results for both time regimes, i.e., short and long time behaviours.
We have proved that at short times the global order parameter we are considering behaves as: 
\begin{equation}
\langle X_c(t)\rangle_\beta=\text{e}^{-\Gamma t}\langle X_c(0)\rangle,
\end{equation}
with $\Gamma=\frac{\Delta}{2}R(\Delta)\r e^{-\Delta\beta}|c|$ and $\Delta=3$ for qutrits. We have also shown that there exits a crossover temperature $T_c$ above which, the initial decay rate for qutrits is smaller than for qubits. Furthermore, we have shown this event only occurs in the case of qutrits, as for other values of $d$, the initial decay rate is always larger than for qubits. On the other hand, far from this initial short-time regime, the topological order of the system decays to zero for times long enough.

\section{Conclusions}
\label{sec:IV}

We have introduced the basic concepts of 2-D Kitaev Model for \emph{qubits} as well as a generalization of the code for \emph{qudits}, i.e. $d-$level systems
with the main purpose of studying its decoherence properties due to thermal effects. 
To this end, we have coupled these systems to thermal baths in order to study the thermal stability within a quantum open systems' formalism, namely  Davies' theory.

The generalization of the toric code leads to new physics. Indeed, we have particularized for the case of \emph{qutrits} and obtained very interesting results. First of all, new abelian anyons have arisen with novel braiding properties, i.e. new statistics by exchange of particles. For instance, let us move a pair of anyons around another pair who stays still. We would pick up a different phase, letting the first pair still and moving the other one around.
Furthermore, new energy processes appear which are forbidden  for \emph{qubits}, being $d=3$ the first non-trivial system where these new processes can be observed. Moreover, we present a master equation that describes the dynamics of any observable of the system coupled to a thermal bath, giving a complete description of the problem.

We have proposed a new way to study thermal stability regarding the loss of topological order in the system. At short times, the system starts loosing its order  with a certain decay rate that we are able to compute explicitly. We have checked that the system relaxes to the thermal state for any value of $d$, as it was expected. However, we have proved that above a certain crossover temperature, the initial decay rate for \emph{qutrits} is smaller than the one from the original case for \emph{qubits}. Surprisingly, this behaviour only happens with \emph{qutrits} and not with other \emph{qudits} with $d>3$.

It would be very interesting to be able to generalize further this study to other topological codes
\cite{TZXLN09,toposubsystem10,noise_topo_subsystem11,YuChenOh07,graphical_nobinary08,rico_briegel08,majorana_codes10} coupled to thermal baths
by deriving appropriate master equations for them. Other challenges in this direction are to study thermal effects with non-abelian topological codes
\cite{MA3,nestedTO08,BS09,BA,BSW11,BSS11,BL11}, higher dimensional codes
\cite{AHHH08,SiYu07,anyonic_loops08,Mandal_Surendran09,Mandal_Surendran10,BLT11,Haah11,Kim11,GTV11,DennisKLP2002,Bacon,Tsomokos} and systems with topological order
based on two-body interactions \cite{Interacting-2body08,Interacting-2body10,SC09,mosaic07},
 instead of many-body interactions in the Hamiltonian. This would facilitate the physical simulation of these topological quantum models \cite{Markus,Muller11,Muller11b,topocorrection09,electronics09,topoJJ08,BAC09,MRLC10}.

\begin{acknowledgments}

We thank the Spanish MICINN grant FIS2009-10061,
CAM research consortium QUITEMAD S2009-ESP-1594, European Commission
PICC: FP7 2007-2013, Grant No.~249958, UCM-BS grant GICC-910758.

\end{acknowledgments}

\appendix

\section{Evolution of the Order Parameter for Qutrits}
\label{App-TOqutrits}

In order to compute $\bra{{\rm GS}}\mathcal{L}(X_c)\ket{{\rm GS}}$ (with $\mathcal{L}(X_c)=\mathcal{L}^x(X_c)+\mathcal{L}^z(X_c)$), we need the expression of the system operators that appear in Eq. (\ref{Operadores3ab}) which were defined previously in Eq. (\ref{opa}) and (\ref{opb}). These operators are expressed in terms of some orthogonal projectors whose definition is given in Eq. (\ref{opP}).
However, there are only two projectors which are relevant here, namely
\begin{equation}
P^j_{++}\ket{{\rm GS}}=\ket{{\rm GS}}\hspace{1cm}\text{and}\hspace{1cm} R^j_{++}\ket{{\rm GS}}=\ket{{\rm GS}},
\end{equation}
as the rest of them vanish when acting on the ground state. Remember that $P^j$ are the projectors associated with the stabilizers $A_s$ and $R^j$ with stabilizers $B_p$.
Moreover we have
\begin{equation}
b_j^{(1)}\ket{{\rm GS}}=0,\hspace{0.3cm}b_j^{(2)}\ket{{\rm GS}}=0,\hspace{0.3cm}b_j^{(2)\dagger}\ket{{\rm GS}}=0,\hspace{0.3cm}b_j^{(0)}\ket{{\rm GS}}=0.
\end{equation}
Thus, after doing some simplifications on Eq. (\ref{Operadores3ab}):
\begin{eqnarray}
 \bra{{\rm GS}}\mathcal{L}_x(X_c)\ket{{\rm GS}}&=&\frac{R(\Delta)}{2}\r e^{-\Delta\beta}\sum_j\bra{{\rm GS}}(2b_j^{(1)}X_cb_j^{(1)\dagger}-b_j^{(1)}b_j^{(1)\dagger}X_c-X_cb_j^{(1)}b_j^{(1)\dagger})\ket{{\rm GS}}=\nonumber\\
 &=&2|c|\bra{{\rm GS}}(\sigma^x_j+(\sigma^x_j)^{-1})X_c\ket{{\rm GS}}=0,
\end{eqnarray}
as $X_c\ket{{\rm GS}}\propto\ket{{\rm GS}}$ but $\sigma^x_j\ket{{\rm GS}}$ is orthogonal to $\ket{{\rm GS}}$, and  we have used the fact that $[P_{\pm,0}^j,X_c]=0$ for every $j$, as these projectors are only functions of vertex operators. This is not true for $R_{\pm,0}^j$ if $j\in c$, i.e. $j$ belongs to the path where $X_c$ is acting on. In that case, since $\sigma^z_j\sigma^x_j(\sigma^z_j)^{-1}=\omega\sigma^x_j$, we obtain $\sigma^z_jX_c(\sigma^z_j)^{-1}=\omega X_c$ for the string operator. In addition, by making use of
\begin{equation}
a_j^{(1)}\ket{{\rm GS}}=0,\hspace{0.3cm}a_j^{(2)}\ket{{\rm GS}}=0,\hspace{0.3cm}a_j^{(2)\dagger}\ket{{\rm GS}}=0,\hspace{0.3cm}a_j^{(0)}\ket{{\rm GS}}=0,
\end{equation}
the result for $\bra{{\rm GS}}\mathcal{L}_z(X_c)\ket{{\rm GS}}$ turns out to be
\begin{eqnarray}
\bra{{\rm GS}}\mathcal{D}_z(X_c)\ket{{\rm GS}}=&=&\frac{R(\Delta)}{2}\r e^{-\Delta\beta}\sum_j\bra{{\rm GS}}(2a_j^{(1)}X_ca_j^{(1)\dagger}-a_j^{(1)}a_j^{(1)\dagger}X_c-X_ca_j^{(1)}a_j^{(1)\dagger})\ket{{\rm GS}}=\nonumber\\
	 &=&\frac{R(3)}{2}\r e^{-3\beta}\sum_j\bra{{\rm GS}}(\sigma^z_j+(\sigma^z_j)^{-1})X_c(\sigma^z_j+(\sigma^z_j)^{-1})\ket{{\rm GS}}-\bra{{\rm GS}}P_{++}^jX_c\ket{{\rm GS}}-\nonumber\\
	 &-&\frac{1}{2}\bra{{\rm GS}}P_{++}^j(\sigma^z_j+(\sigma^z_j)^{-1})P_{++}^jX_c\ket{{\rm GS}}-\bra{{\rm GS}}X_cP_{++}^j\ket{{\rm GS}}-\nonumber\\
	&-&\bra{{\rm GS}}X_cP_{++}^j(\sigma^z_j+(\sigma^z_j)^{-1})P_{++}^j\ket{{\rm GS}}=\nonumber\\
	&=&\frac{R(3)}{2}\r e^{-3\beta}\sum_j\delta_{j\not\in c}(\bra{{\rm GS}}(2+\sigma^z_j+(\sigma^z_j)^{-1})X_c\ket{{\rm GS}}-\bra{{\rm GS}}X_c\ket{{\rm GS}}-\nonumber\\
	 &-&\frac{1}{2}\bra{{\rm GS}}(\sigma^z_j+(\sigma^z_j)^{-1})X_c\ket{{\rm GS}}-\bra{{\rm GS}}X_c\ket{{\rm GS}}-\frac{1}{2}\bra{{\rm GS}}(\sigma^z_j+(\sigma^z_j)^{-1})X_c\ket{{\rm GS}})+\nonumber\\
	&+&\delta_{j\in c}(\bra{{\rm GS}}(\sigma^z_j+(\sigma^z_j)^{-1})(\omega^2\sigma^z_j+\omega(\sigma^z_j)^{-1})X_c\ket{{\rm GS}}-2\bra{{\rm GS}}X_c\ket{{\rm GS}}-\nonumber\\
	&-&\bra{{\rm GS}}(\sigma^z_j+(\sigma^z_j)^{-1})X_c\ket{{\rm GS}})=-\frac{3}{2}R(3)\r e^{-3\beta}|c|\bra{{\rm GS}}X_c\ket{{\rm GS}}=\nonumber\\
&=&-\frac{\Delta}{2}R(\Delta)\r e^{-\Delta\beta}|c|\bra{{\rm GS}}X_c\ket{{\rm GS}},\nonumber
\end{eqnarray}
	
where $|c|$ is the number of points in the path $c$.

\section{Irreducibility of the Computational Representation of the $d$--Pauli Group}
\label{App-irreducibility}

The $d$--Pauli group is generated by products of $\sigma_x$ and $\sigma_y$ such that $\sigma_x^d=\sigma_z^d=\mathds{1}$ and $\sigma_z\sigma_x=\omega\sigma_x\sigma_z$ where $\omega$ is a primitive $d$--root of the unity. Its order is $d^3$, which is a direct consequence that any element of the group can be written as $\omega^n\sigma_x^m\sigma_z^k$ for some $n$, $m$ and $k$.

We take the representation of the $d-$Pauli group when acting on the computational basis:
\begin{eqnarray}
\sigma_x|n\rangle&=&|n+1\rangle \quad \mathrm{mod.}\ d,\\
\sigma_z|n\rangle&=&\omega^n|n\rangle,
\end{eqnarray}
and we want to show this representation is irreducible. We proceed by computing the character $\chi$ of every of its elements, which is given by the trace of the matrices. Using the computational basis when taking the trace, from the above relations, $\chi(\sigma_x^m)=0$ for $m\in\{1,\ldots,d-1\}$. Similarly $\chi(\sigma_z^m)=0$ for $m\in\{1,\ldots,d-1\}$ as the sum of the roots of the unity vanishes. On the other hand, because $\sigma_z\sigma_x=\omega\sigma_x\sigma_z$ and the cyclic property of the trace, we conclude that the character of every element of the form $\sigma_x^m\sigma_z^k$ is zero for any representation. The rest of the terms are proportional to the identity $\omega^n\mathds{1}$, and so $\chi(\omega^n\mathds{1})=\omega^nd$.

The irreducibility criterium asserts \cite{JansenBoon,NC} that a representation of a group $G$ is irreducible if and only if the scalar product of characters is the identity, this is
\begin{equation}
(\bm{\chi},\bm{\chi})=\frac{1}{|G|}\sum_{g\in G}\chi^\ast(g)\chi(g)=1,
\end{equation}
where $|G|$ is the order of the group. For the computational representation of the $d-$Pauli group we have
\begin{equation}
(\bm{\chi},\bm{\chi})=\frac{1}{d^3}\sum_{n=0}^{d-1}(\omega^n d)^\ast \omega^n d=\frac{1}{d}\sum_{n=0}^{d-1}|\omega|^n=1,
\end{equation}
thus,  the representation is irreducible.


\begin{thebibliography}{99}
\bibitem{notunique} Indeed, there are different generalizations for the operators $X$ and $Z$. What makes simple the generalization of the toric code to higher dimensions is to keep the action of $X$ and $Z$ on the computational basis to be analogous to the case of qubits. This implies a specific structure for the anticommutation rule, namely $XZ=\omega ZX$, where $\omega$ is a primitive $d-$root of the unity. Note for instance that, another common generalization of $X$ and $Z$, based on the generators of the Lie algebra $\mathfrak{su}(d)$, does not fulfill these anticommutation relations.
\end{thebibliography}
\end{document}